\documentclass[10pt]{article}  

\usepackage{amsmath,graphicx,bbm,amssymb}
\usepackage{caption}
\usepackage{color}
\usepackage{tikz}
\usetikzlibrary{shapes.multipart}
\usetikzlibrary{shapes.misc, positioning}
\usepackage{fancyhdr}
\usepackage{colortbl}
\usepackage{marvosym}
\usepackage[hmargin=1.5cm, top=2cm, bottom=2cm]{geometry}
\usepackage[colorlinks=true,linkcolor=blue,citecolor=blue,urlcolor=blue]{hyperref}
\usepackage[export]{adjustbox}
\usetikzlibrary{arrows} 
\usepackage{titling}
\usepackage[affil-it]{authblk}

\newcommand{\norme}[1]{\left\vert\left\vert #1 \right\vert\right\vert}
\newcommand{\para}[1]{\left(#1\right)}
\newcommand{\cro}[1]{\left[#1\right]}
\newcommand{\aver}[1]{\left\langle #1 \right\rangle}

\newcommand{\rz}{r_0}
\newcommand{\lz}{L_0}
\newcommand{\cnh}{C_n^2(h)}

\newcommand{\otf}[1]{\text{OTF}_{#1}}

\newcommand{\cov}[1]{\mathcal{C}_{#1}}
\newcommand{\rhovec}{\boldsymbol{\rho}}
\newcommand{\rvec}{{\mathbf{r}}} 
\newcommand{\Pup}{{\mathcal{P}}} 

\newcommand{\wOTF}{\widehat{\text{OTF}}_\varepsilon}

\newcolumntype{P}[1]{>{\centering\arraybackslash}p{#1}}
\newcommand{\gDM}{g_\text{DM}}
\newcommand{\gTT}{g_\text{TT}}
\newcommand{\az}{\boldsymbol{a}_\text{z}}
\newcommand{\rhol}{\boldsymbol{\rho}/\lambda}

\title{\LARGE PRIME: Psf Reconstruction and Identification for Multiple sources characterization Enhancement. Application to Keck NIRC2 imager.}

\author{
	O. Beltramo-Martin,$^{1,2}$\thanks{E-mail: olivier.beltramo-martin@lam.fr}
	C.M. Correia,$^{1}$	
	S. Ragland,$^{3}$
	L. Jolissaint,$^{4}$
	B. Neichel,$^{1}$        
	T. Fusco,$^{1,2}$	 		
	P.L. Wizinowich$^{3}$   
	}
	
	\affil{
	$^{1}$Aix Marseille Univ., CNRS, CNES LAM, 38 rue F. Joliot-Curie, 13388 Marseille, France\\	
	$^{2}$ONERA, The French Aerospace Laboratory  BP. 72, F-92322 Chatillon Cedex, France\\
	$^{3}$W. M. Keck Observatory, 65-1120 Mamalahoa Hwy, Kamuela, HI 96743\\
	$^{4}$University of Applied Sciences Western Switzerland, 1401 Yverdon-les-Bains, Switzerland	
}

\date{}

\begin{document} 
	\maketitle
	\begin{abstract}
		
	In order to enhance accuracy of astrophysical estimates obtained on Adaptive-optics (AO) images, such as photometry and astrometry, we investigate a new concept to constrain the Point Spread Function (PSF) model called PSF Reconstruction and Identification for Multi-sources characterization Enhancement (PRIME), that handles jointly the science image and the AO control loop data. We present in this paper the concept of PRIME and validate it on Keck II telescope NIRC2 images. We show that by calibrating the PSF model over the scientific image, PSF reconstruction achieves 1\% and 3 mas of accuracy on respectively the Strehl-ratio and the PSF full width at half maximum. We show on NIRC2 binary images that PRIME is sufficiently robust to noise to retain photometry and astrometry precision below 0.005 mag and 100$\mu$as on a $m_H=$ 14 mag object. Finally, we also validate that PRIME performs a PSF calibration on the triple system Gl569BAB which provides a separation of 66.73$\pm 1.02$ and a differential photometry of 0.538$\pm 0.048$, compared to the reference values obtained with the extracted PSF which are 66.76 mas $\pm$ 0.94 and 0.532 mag $\pm$ 0.041.
	  
	\end{abstract}
	
	
\section{Introduction}
\label{S:Intro}
	
Estimation of key science quantities, such as astrometry and photometry are made possible by data-reduction techniques. For sources extraction or modeling, common pipelines are StarFinder (\cite{Diolaiti2000}), SeXTRACTOR (\cite{Bertin1996}), or DAOPHOT (\cite{Stetson1987}). Nonetheless, past analysis (\cite{Fritz2010,Yelda2010,Shodel2010,Sheehy2006}) have assessed that 50\% of the astrometry and photometry error breakdown in the Galactic center (GC) is induced by a problem of determination of the Point spread function (PSF) model. A more recent study (\cite{Ascenso2015}) has illustrated that photometry in the general case of globular clusters observations assisted by Adaptive Optics (AO) can be obtained at better than 1\%-level by providing StarFinder with the exact PSF, advocating for pushing the PSF characterization further to reach more accurate science estimates. One may rapidly realize that the source confusion and measurement noise are important limitations to extract the PSF.\\

An alternative approach to identifying the AO PSF is PSF reconstruction (PSF-R) (\cite{Veran1997}) which estimates the PSF from real-time AO control loop data, i.e. wavefront sensor (WFS) measurements and deformable mirrors (DM) commands. Such a concept has already been demonstrated on-sky (\cite{Ragland2018_PSFR,Martin2016JATIS,Ragland2016,Jolissaint2015,Veran1997}) 
and gives the astronomer access to a PSF model without any extraction from the focal-plane, which has the great advantage of achieving a PSF determination insensible to science detector noise, potential source confusion or absence of PSF calibrators within the science field. However, PSF-R lacks operational use so far because it requires a large amount of specific AO skills, especially to capture comprehensively AO performance and behavior on-sky. For instance, in the case of AO-assisted observations led at the W.M. Keck Observatory with a quad-cells WFS, the determination of WFS optical gains plays a major role in the PSF reconstruction and is highly sensitive to seeing conditions and may vary unexpectedly (\cite{Ragland2016}).

Moreover, the AO telemetry permits to reconstruct the PSF in the Guide Star (GS) direction only. To extrapolate the PSF across the Field of View (FOV), one must model the spatial variations due to anisoplanatism (\cite{BeltramoMartin2018_aniso,Fried1982}), whose model relies on atmospheric parameters such as the $\cnh$. For single-conjugated AO (SCAO) systems that rely on a single Natural Guide Star (NGS) or a  Laser Guide Star (LGS) plus a NGS to measure tip-tilt modes, the anisoplanatism model can not be calibrated from AO control loop data, which necessitates to obtain the $\cnh$ from external instruments (\cite{Osborn2018,Butterley2006,Wilson2002}) pointing at different sky directions than the telescope. Discrepancies at a level of 10\% up to 20\% on the $\cnh$ estimation compared to AO control loop data-based technique on multiple NGS/LGS systems has been already observed (\cite{Ono2016}), which has an impact of the off-axis PSF model accuracy.

On top of that, the PSF is affected by field-dependent static aberrations that must be characterized by using long-exposure phase diversity (\cite{Mugnier2008}) or focal-plane sharpening (\cite{Lamb2016}) for instance, which potentially requires a large amount of telescope time (\cite{Sitarski2014}). Finally, earlier studies have showed discrepancies between the reconstructed PSFs and the on-sky PSFs (\cite{Ragland2016,Jolissaint2015}), that are attributed to the presence of instrumental effects, addressed at the system level (\cite{Ragland2018_PSFR,Ragland2018_COPHASING}) and telemetry issues.\\

In this paper, we present PRIME, which stands for PSF Reconstruction and Identification for Multiple sources characterization Enhancement, as a supplementary methodology to PSF-R that addresses the discrepancies in the post-processing stage. PRIME aims to calibrating a PSF model, preliminary instantiated from AO control loop data, by adjusting some key atmosphere and system parameters, such as the seeing, the $\cnh$ and WFS optical gains, over PSF extracted out from the focal-plane images. The calibration process is made from a best-fitting process based on a non-linear least-squared minimization. When this calibration done, one may extrapolate the PSF model at any science sky directions or wavelengths.

PRIME gathers all relevant information to identify the PSF model as accurately as possible. It benefits the PSF-R framework to build a rich an accurate PSF model and tackle mis-identification of key parameters by retrieving them from the focal-plane image. PRIME has however the disadvantage to require the presence of one or several PSFs within the observed field, which suits only to a limited number of science cases. Therefore, our strategy is to deploy PRIME on engineering data and observations of globular clusters to learn how the parameters evolve with respect to observing conditions, such as the seeing or wind-shake for instance, and infer the PSF model without additional focal-plane calibration. In the future, we will also investigate for coupling the PSF-R facility at the Keck telescopes with the AIROPA package (\cite{Ciurlo2018,Witzel2016}) to enhance scientific exploitation of the near infra-red imager NIRC2 and the integral field spectrograph OSIRIS (\cite{Larkin2006,McLean2000}).

In Sect.~\ref{S:PRIME}, we detail the PRIME framework that relies on the PSF-R model proposed by (\cite{Jolissaint2015}). We highlight the key model parameters and validate the approach performance in simulation and analyze its sensibility with respect to the star magnitude. We present a model validation over KECK II telescopes NIRC2 images in Sect.~\ref{S:NIRC2}. We validate that the approach delivers PSF-based estimates, such as Strehl-ratio (SR) and Full Width at Half Maximum (FWHM), at 1\%-level and compare the model outputs to external references. Sect.~\ref{S:binaries} illustrates an application of PRIME on tight-binaries characterization to asses accuracy and precision of astrometry and photometry.

\section{PSF calibration with PRIME}
\label{S:PRIME}	

\subsection{PSF direct model}
\label{SS:theory}	
	
Classical PSF-R allows to model the system Optical transfer function~(OTF) in closed loop from AO control loop data. The robustness and efficiency of this method is well established in the literature (\cite{Ragland2018_PSFR,Martin2016JATIS,Jolissaint2015,Gilles2012,Flicker2008,Veran1997}). Our analysis focuses on the Keck SCAO system guided by either a NGS or LGS and relies on the PSF-R concept implemented by \cite{Jolissaint2015}. We have identified a small number of key parameters whose fitting over sky images makes the reconstruction more accurate. We define $\boldsymbol{\mu}$ the meta-parameters vector that gathers the model parameters, which are $\cnh$ the atmospheric profile, $\gDM$ and $\gTT$ the WFS and Tip-Tilt (TT) optical gains and $\az$ the list of Zernike modes coefficients that allow to include an additional static phase. The model is passed to an iterative non-linear minimization algorithm that adjusts the parameters to match the observed PSF, as detailed in Sect.~\ref{SS:PRIME}. The model expands as follows
\begin{equation}
\wOTF(\rhol,\boldsymbol{\mu}) =  \otf{\text{Stat}}(\rhol).\text{K}_{\perp}(\rhol,\rz).\text{K}_{\text{TT}}(\gTT).\text{K}_{\text{DM}}(\rhol,\cnh,\gDM).\text{K}_z(\rhol,\az)	
\label{E:otfdot}
\end{equation}
where 
\begin{itemize}
\item[$\bullet$] $\otf{\text{Stat}}$ includes the pupil model and the residual static aberrations, such as Non common path aberrations (NCPA), telescope-dependent static-aberrations and segmented mirror cophasing errors. We assume in this paper that the model is fed by the calibration of those static terms using dedicated phase diversity (\cite{Mugnier2008}) or focal sharpening (\cite{Lamb2016}) methods. For Keck observations, the calibration procedure is detailed in (\cite{Ragland2016,Witzel2016,Sitarski2014}). If $\varphi_\text{Stat}$ is the static phase of the electric-field in the pupil during the observation, $\otf{\text{Stat}}$ is given by
\begin{equation}
\label{E:otfTel}
\otf{\text{Stat}}(\rhol) =
\iint_\Pup \Pup(\rvec)\Pup^{*}(\rvec + \rhovec).\exp\para{i(\varphi_\text{Stat}(\rvec) + \varphi_\text{Stat}(\rvec + \rhovec)) }d\rvec
\end{equation}
where $\Pup$ is the pupil function on which the static phase is defined, $\rvec$ the phase sample coordinates inside the pupil, $\rho$ is the separation vector within the pupil and $\lambda$ the wavelength. Calibration of static terms during the observations we have reduced were done right before the image acquisition.\\

\item[$\bullet$]  $\text{K}_{\perp}$ is a spatial filter that includes both the fitting and WFS aliasing error by following expressions given in (\cite{Correia2014}). $\text{K}_{\perp}$ is derived from the atmospheric phase Power spectrum density~(PSD) that scales with $\rz^{-5/3}$. \\

\item[$\bullet$] $\text{K}_{\text{TT}}$ is the TT spatial filter derived from the residual AO jitter. We name $w_\text{TT}(t)$ the time-dependent residual TT wavefront delivered by the tip-tilt sensor. $\otf{\text{TT}}$ is readily obtained from the tip-tilt phase Structure function~(SF) $\mathcal{D}_\text{TT}$ as
\begin{equation}
	\text{K}_{\text{TT}}(\rhol,\gTT) = \exp\para{-0.5\times \mathcal{D}_\text{TT}(\rhovec,\gTT)},
\end{equation}	
where $\mathcal{D}_\text{TT}(\rhovec)$ is derived from $w_\text{TT}$ based on
\begin{equation}
	\mathcal{D}_\text{TT}(\rhovec,\gTT) = \dfrac{g^2_\text{TT}}{D^2}\times\cro{\rhovec.\para{\aver{w_\text{TT}(t).w_\text{TT}(t)^t}_t - \cov{\eta}^\text{TT}}.\rhovec^t}
\end{equation}
where $\aver{x}_t$ is the temporal average of vector $x$, $\cov{\eta}^\text{TT}$ the tip-tilt noise covariance matrix and $D$ the telescope diameter. The scalar $\gTT$ is a tuning factor of the TT optical gain. Noise variance is obtained by adjusting a model of the AO loop rejection filter on temporal power spectrum density as proposed by \cite{Jolissaint2015}.\\

\item[$\bullet$] $\text{K}_{\text{DM}}$ corresponds to the TT excluded filter calculated from the AO residual error within the AO control radius. It is calculated from $\mathcal{D}_\theta(\rvec,\rhovec)$ the residual phase SF in the science direction $\theta$ as follows
\begin{equation}
\label{E:otfDM}
\text{K}_{\text{DM}}(\rhol,\cnh,\gDM) =
\dfrac{\iint_\Pup  \Pup(\rvec)\Pup^*(\rvec + \rhovec) \exp\para{-0.5\times \mathcal{D}_\theta(\rvec,\rhovec,\cnh,\gDM)} d\rvec }{\iint_\Pup  \Pup(\rvec)\Pup^*(\rvec + \rhovec) d\rvec}
\end{equation}
where $D_\theta$ expands into two terms
\begin{equation}
	\mathcal{D}_\theta(\rvec,\rhovec) = \mathcal{D}_0(\rvec,\rhovec,\gDM) + \mathcal{D}_\Delta(\rvec,\rhovec,\cnh),
\end{equation}
with $\mathcal{D}_0$ as the TT excluded residual phase SF in the guide star direction and $\mathcal{D}_\Delta$ the anisoplanatism SF. A complete methodology to derive $\mathcal{D}_\Delta$ from the $\cnh$ is given in (\cite{BeltramoMartin2018_aniso}).
$\mathcal{D}_0$ is derived from 
\begin{equation}
\mathcal{D}_0(\rvec,\rhovec,\gDM) =g^2_\text{DM}\times\textbf{F}(\rvec).\para{\aver{w_\text{DM}(t).w_\text{DM}(t)^t}_t - \cov{\eta}^\text{DM}}.\textbf{F}^t(\rvec+\rhovec),
	\label{E:cov0}
\end{equation}
where $w_\text{DM}$ is the time-dependent residual wavefront in the DM commands space, $\cov{\eta}^\text{DM}$ is the noise covariance matrix on DM actuators, $\textbf{F}$ the influence function matrix that converts the DM commands into a zonal description of the phase. Again, the scalar $g_\text{DM}$ is a tuning factor of the WFS optical gain.\\

\item[$\bullet$] $\text{K}_z$ permits to incorporate $n_z$ additional static modes, presently Zernike modes, into the PSF model
\begin{equation}
\label{E:otfStat}
\text{K}_{z}(\rhol,\az) =\dfrac{\iint_\Pup  \Pup(\rvec)\Pup^*(\rvec + \rhovec) \exp\para{i(\varphi_\text{z}(\rvec) + \varphi_\text{z}(\rvec + \rhovec)}d\rvec }{\iint_\Pup  \Pup(\rvec)\Pup^*(\rvec + \rhovec) d\rvec}, 
\end{equation}
with
\begin{equation}
\varphi_\text{z}(\rvec) = \sum_i^{n_z} \az.Z_i(\rvec).
\end{equation}
\end{itemize}

Including additional Zernike modes, like focus and astigmatism terms, will be particularly useful to compensate for any static terms that are not well identified. With the current formulation, we have the ability to describe an exact PSF for a given wavelength and field position with a minimal number of parameters: 3 for on-axis reconstruction (seeing, optical gains) up to 9 for off-axis reconstruction with a description of the $\cnh$ profile over 7 layers. In the following, we will limit the number of reconstructed Zernike modes to focus and astigmatism terms only.
 
\subsection{Problem fitting}
\label{SS:PRIME}

\begin{figure*}
	\centering		
	\includegraphics[width=17cm]{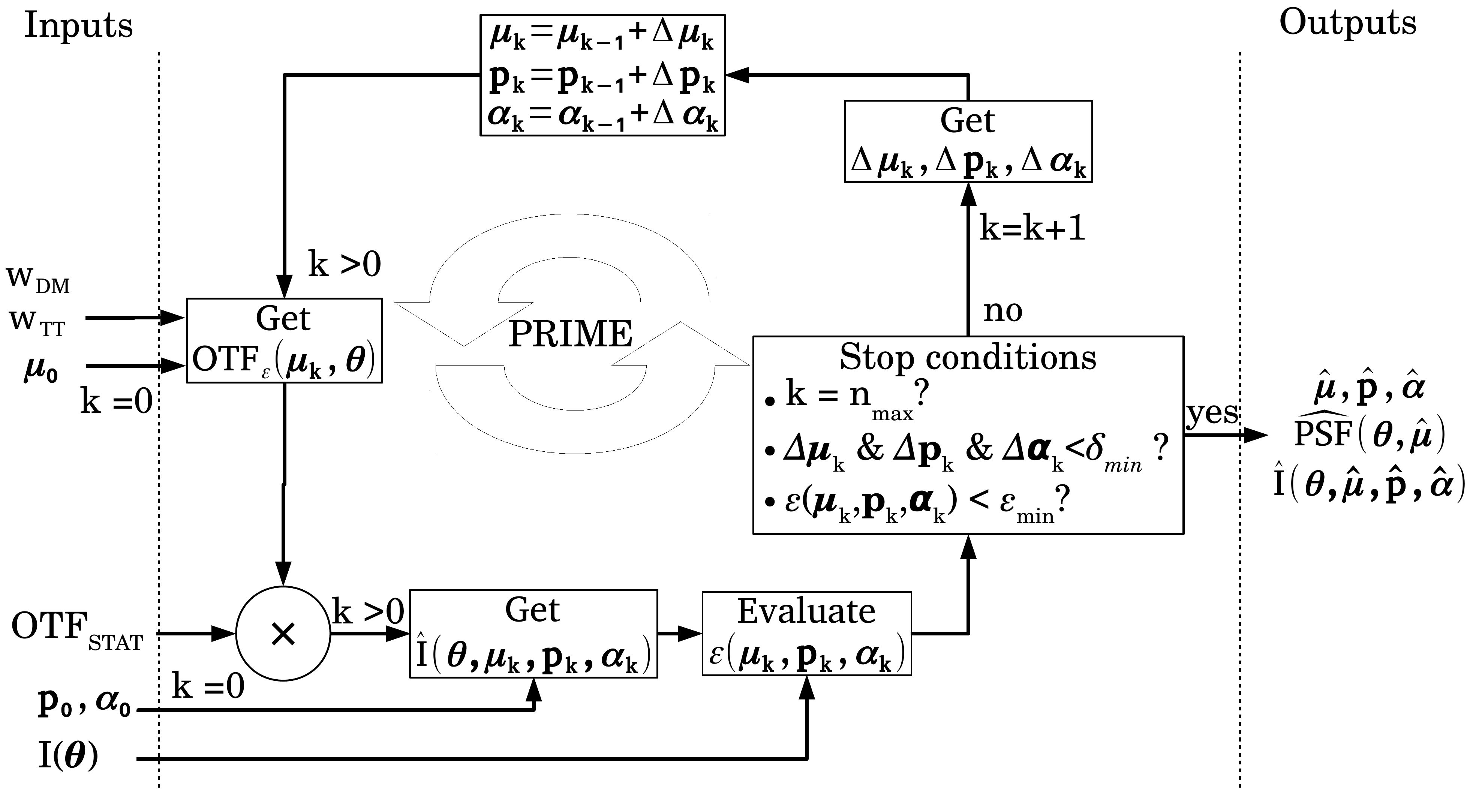}
	\caption{Block-diagram of PRIME algorithm. Increments on inputs are derived from image model gradient by the Levenberg-Marquardt algorithm. In the end of the procedure, we get the estimated PSF parameters vector $\widehat{\boldsymbol{\mu}}$, the stellar photometry $\widehat{\boldsymbol{p}}$ and astrometry $\widehat{\boldsymbol{\alpha}}$, the corresponding PSF and image model.}
	\label{F:PRIME}
\end{figure*}

We have deployed an iterative non-linear minimization to best-fit the PSF model over observed images. We present the PRIME architecture in Fig.~\ref{F:PRIME}. We remind that $\boldsymbol{\mu} = [\cnh,g_\text{DM},g_\text{TT},\az]$ is a PSF meta-parameters that PRIME retrieves by minimizing the the following multiple point-sources criterion
\begin{equation}
	\begin{aligned}
		&\varepsilon(\boldsymbol{\mu},\boldsymbol{p},\boldsymbol{\alpha}) =
		&\sum_{i=1}^{n_\text{psf}} \norme{\para{I(\theta_i) - \widehat{I}(\theta_i,\boldsymbol{\mu},p_i,\boldsymbol{\alpha}_i) + b}}^2_{\mathcal{L}_2(\Omega)},
	\end{aligned}
	\label{E:criteria}
\end{equation}
where $\text{I}(\theta_i)$ is the observed image at the sky location $\theta_i$ and $\widehat{I}(\theta_i,\boldsymbol{\mu},p_i,\boldsymbol{\alpha}_i)$ is the corresponding model that depends on PSF parameters $\boldsymbol{\mu} = [\cnh,g_\text{DM},g_\text{TT}]$, stellar photometry $p_i$, sub-pixel astrometry $\boldsymbol{\alpha}_i$ around position $\theta_i$ and $b$ a scalar value to adjust the background level. We use a $\mathcal{L}_2$ norm that integrates the residual over pixels within a square box of field of view given by $\Omega$. We calculate the norm over at least twice the AO correction radius.

The image model is then given by the following expression
\begin{equation}
	\begin{aligned}
		\widehat{I}(\theta_i,\boldsymbol{\mu},p_i,\boldsymbol{\alpha}_i) = p_i\times \mathcal{F}_{\Omega}\cro{\wOTF(\theta_i,\boldsymbol{\mu})\times \exp^{-2i\pi\times\boldsymbol{\alpha}_i.\boldsymbol{\rho}/\lambda}}
	\end{aligned}
	\label{E:image}
\end{equation}
where $\wOTF$ is the model OTF given by Eq.~\ref{E:otfdot} that depends on the meta-parameters $\mu$ and is multiplied by the exponential term to introduce sub-pixel shifts. The operator $\mathcal{F}_\Omega$ applies the 2D Fourier transform and any required operations (zero-padding, cropping, interpolation) to get the model PSF at the desired resolution, position and pixel-scale. Practically, we compute the PSF model over $n\times\Omega$ pixels, with $1<n<2$ to capture the highest spatial frequency in the box. For a sole PSF, the PSF is centered and having $n=1$ is enough, but when having multiple sources in the field, it can be necessary to get to $n=2$ if stars are at image corners.\\

PRIME solves for this criterion iteratively using a Levenberg-Marquardt algorithm which derives at each iteration the PSF model gradient with respect to the adjusted parameters. The minimization process stops when reaching the stop conditions, which can be either a maximal number of iterations, or the variations on adjusted parameters or criterion are lower than a threshold given by the user. In practice, we have found that $n_\text{max} = 300$, $\delta_\text{min}=10^{-10}$ and $\varepsilon_\text{min} = 10^{-10}$ provides a satisfactory convergence for Keck II telescope NIRC2 images. Eventually, PRIME delivers astrometry and photometry of stellar sources, the post-AO PSF best-fitted model and the corresponding parameters. So far, the method is only available to work over a limited number of stars (20) for computing time reasons and must be fed with initial stellar astrometry at few pixels accuracy.\\

Also, Eq.~\ref{E:criteria} does include a data-based criteria only and no regularization term. To avoid the algorithm convergence to a local minimum, we pass to PRIME initial inputs that are assumed to be close to what expected, such as the MASS/DIMM $\cnh$ and the optical gain calibration (\cite{Ragland2016}) and we bound accordingly the parameters space. By applying PRIME over more and more archive data, we will multiply the feedback on the system and make its performance more stable and predictable, as well as sampling the probability density function of the key parameters. Consequently, PRIME is going to be updated with respect to the what we will learn on future tests to make it as  more robust and efficient as possible.

\subsection{Image-assisted wavefront error breakdown}
\label{SS:wfe}
The wavefront error budget is calculated from the trace of covariance matrices that feed the PSF model described in Sect.~\ref{SS:theory}. Thanks to the PSF calibration achieved by PRIME, the covariance matrix of the residual DM-compensated phase and tip-tilt are naturally scaled properly regarding the PSF through optical gains tuning factors $\gDM$ and $\gTT$. This occurs similarly for the DM-fitting and anisoplanatism errors thanks to the atmospheric parameters adjustment process. The use of the sky PSF adds a supplementary constrain on the absolute level of each term that must deliver a overall Strehl-ratio (SR) very close to the image SR. 

The total wavefront error is given by
\begin{equation}
\label{E:wfe}
\sigma^2_\varepsilon = \sigma^2_\text{stat} + \sigma^2_\text{fit} + \sigma^2_\text{dm} + \sigma^2_\text{jitter} + \sigma^2_\text{meas} + \sigma^2_\text{ang-aniso} + \sigma^2_\text{foc-aniso} + \sigma^2_\text{tilt-aniso} 
\end{equation}
where
\begin{itemize}
\item[$\bullet$]  $\sigma^2_\text{stat}$ refers to the static aberrations error that contains NCPA calibration residual, telescope co-phasing error if calibration available and additional Zernike terms adjusted by PRIME. It is derived from the variance of the static phase map $\varphi_\text{stat} + \varphi_z$ over the pupil.\\

\item[$\bullet$] $\sigma^2_\text{fit}$ is the fitting error that corresponds to the uncompensated high-spatial frequencies and derived from the adjusted seeing estimation from PRIME by integrating the fitting PSD (\cite{Jolissaint2010}).\\

\item[$\bullet$] $\sigma^2_\text{dm}$ is the DM bandwidth error that computes from the trace of the matrix $\widehat{g}^2_\text{DM}\times\aver{w_\text{DM}(t).w^t_\text{DM}(t)} - \cov{\eta}^\text{DM}$ that appears in the calculation of $\otf{\text{DM}}$ in Eq.~\ref{E:otfDM}. It gives the AO correction residual excluded from high-spatial frequencies, tip-tilt modes and measurements errors, which mostly corresponds to the servo-lag error.\\

\item[$\bullet$] $\sigma^2_\text{jitter}$ is the jitter error given by the trace of the matrix $\widehat{g}^2_\text{TT}\times\aver{w_\text{TT}(t).w^t_\text{TT}(t)} - \cov{\eta}^\text{TT}$ that gather the residual tip-tilt error.\\

\item[$\bullet$] $\sigma^2_\text{meas}$ corresponds to the measurement error that includes the WFS noise and aliasing filtered by the AO loop. The WFS noise error is estimated from the telemetry (\cite{Jolissaint2015}) and the aliasing error is given by integrated the aliasing PSD across spatial frequencies (\cite{Jolissaint2010}) with the adjusted seeing value.\\

\item[$\bullet$] $\sigma^2_\text{ang-aniso},\sigma^2_\text{foc-aniso}$ and $\sigma^2_\text{tilt-aniso}$ refer respectively to the angular (off-axis operations), focal and tilt (LGS operations) anisoplanatism errors. For the present analysis, we have only considered focal anisoplanatism for LGS cases regarding the on-axis position of the NGS. Errors are derived from the trace of covariance matrices associated to each of these three terms (\cite{BeltramoMartin2018_aniso}) after adjustment of $\boldsymbol{\mu}$ with PRIME.
\end{itemize}
	 	
\subsection{Performance in simulation}

\begin{figure*}
	\centering
	\includegraphics[width=5.8cm]{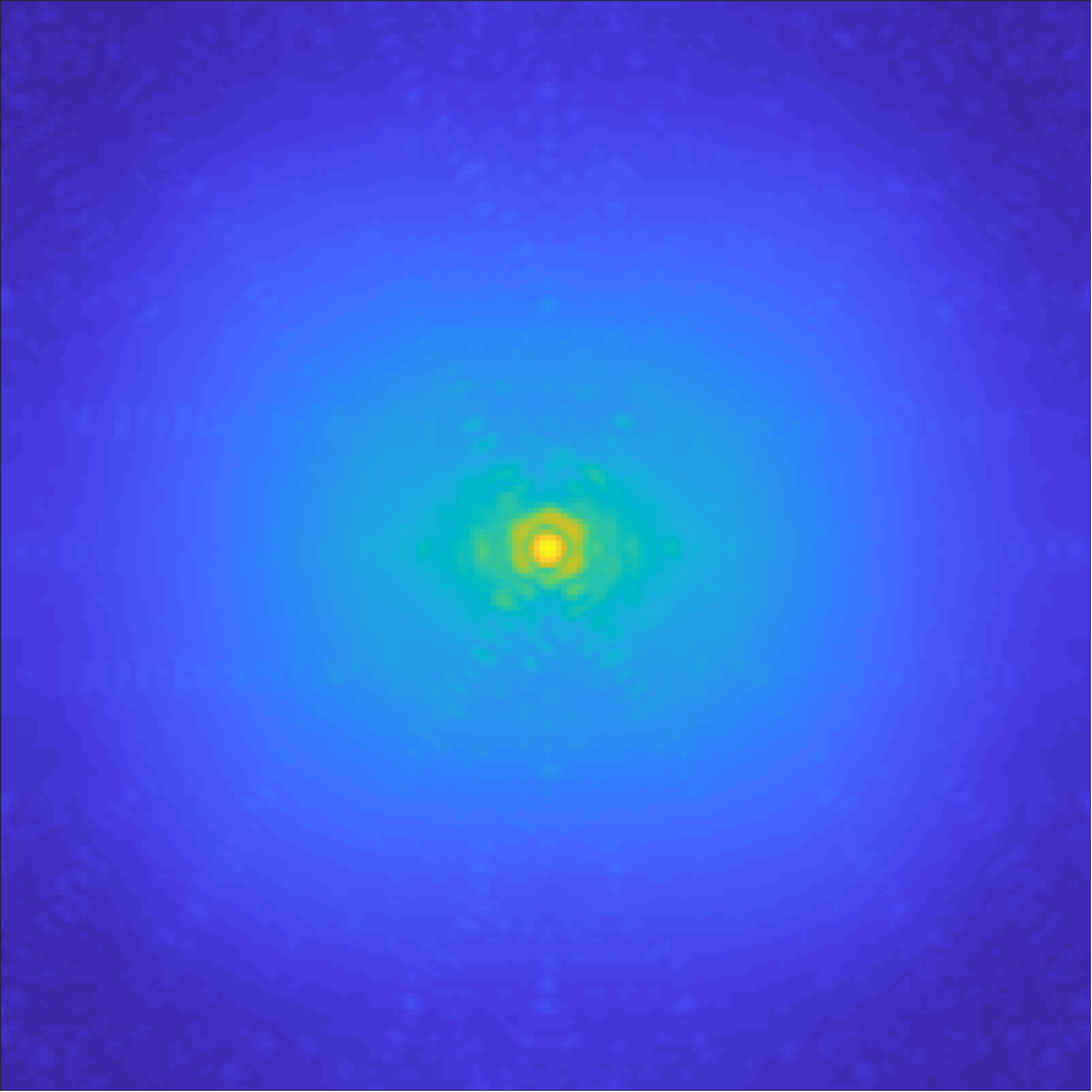}
	\includegraphics[width=5.8cm]{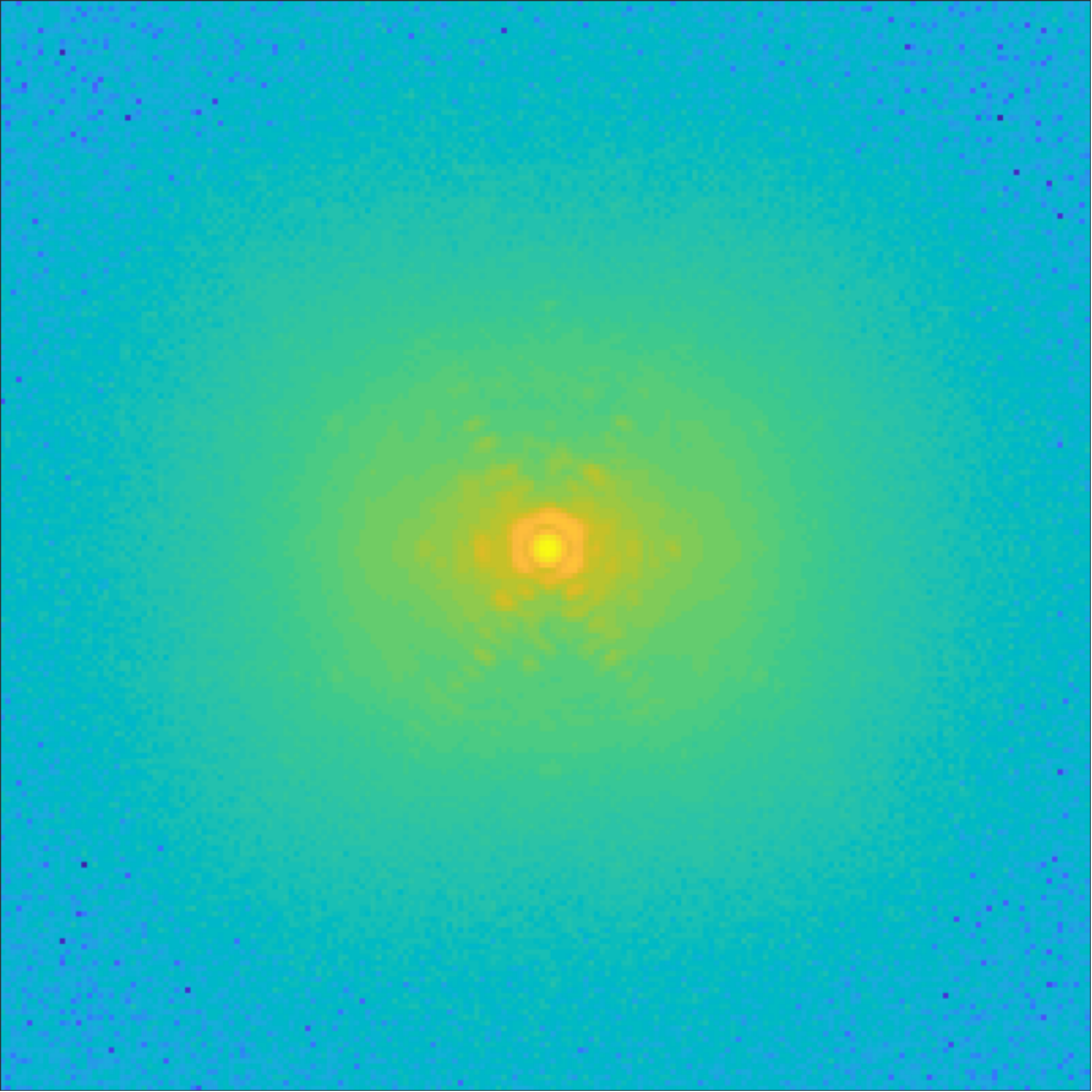}
	\includegraphics[width=5.8cm]{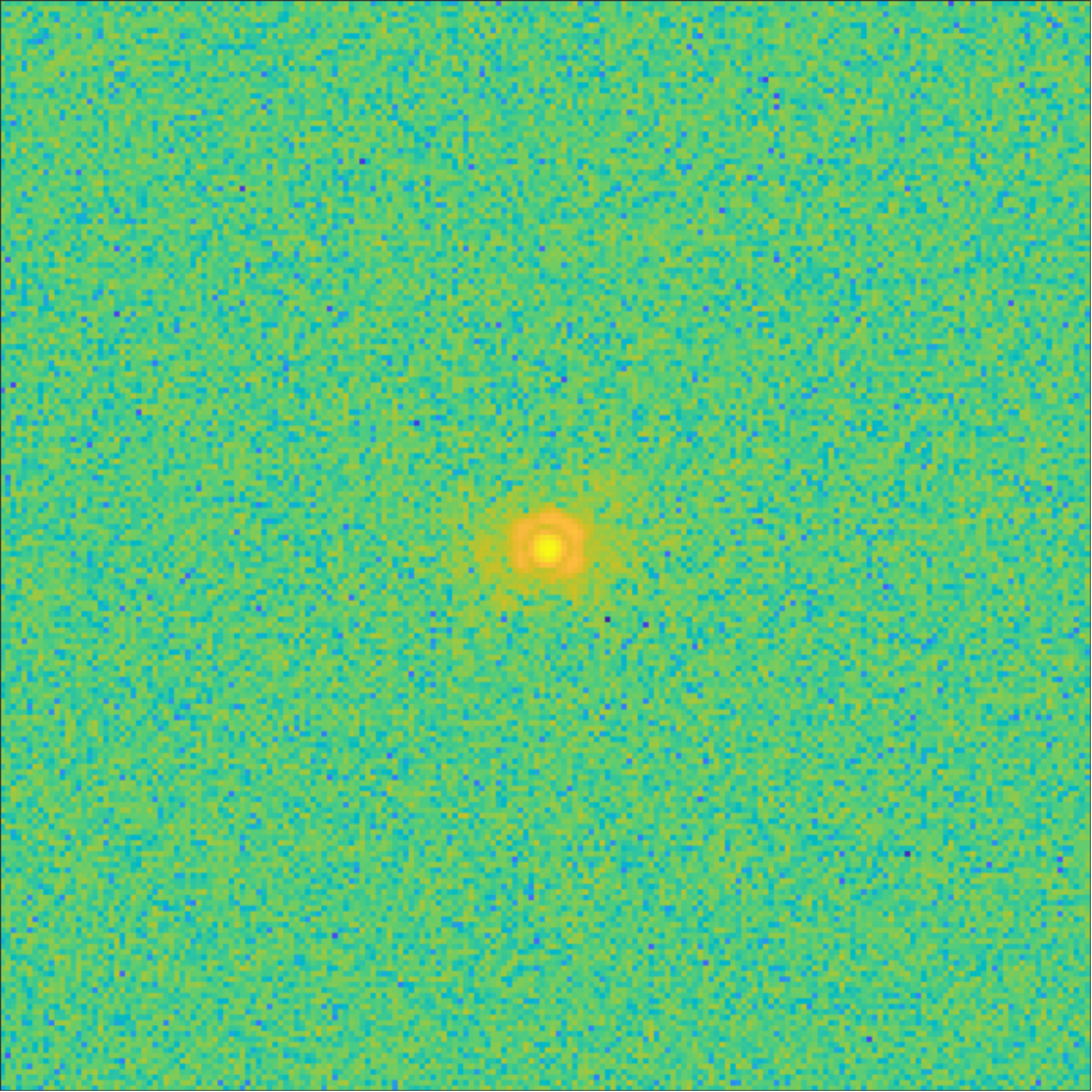}
	\caption{Simulated NIRC2 PSFs in H-band narrow field mode with 30s of exposure time including electronic and photon noise. The PSF is affected by NCPA residual and is located at 10 arcsec from the guide star. The isoplanatic angle was set to $\theta_0=25$ arcsec and $\rz=16$cm. \textbf{Left:} No noise \textbf{Middle:} magnitude 9 \textbf{Right:} magnitude 14.}
	\label{F:psfSimu}
\end{figure*}

We have simulated a AO-compensated H-band PSFs illustrated in Fig.~\ref{F:psfSimu} observed by NIRC2 and located at 10 arcsec from the guide star, which corresponds to the maximal separation between two stars observed with the NIRC2 detector in narrow field mode with 9.94 mas/pixel. We have contaminated the PSF with star photon-noise, electronic noise with 38 photo-events of standard-deviation, 0.08 e-/s of dark current and 13.6 mag/arcsec of sky brightness.

The simulation includes NCPA residual and anisoplanatism, with a 2-layers atmospheric profile (64\% at 180m and 36\% at 7km) with $\rz=16$cm (500nm) and $\lz$=25m. The simulated profile is defined by compressing with the mean-weight approach (\cite{Robert2010}) the median Maunakea profile over two layers, which is sufficient to describe the anisoplanatism at this separation (\cite{BeltramoMartin2018_aniso}). Fig.~\ref{F:psfSimu} shows a slight PSF elongation due to anisoplanatism, which is drowned into the noise floor when including detector noise.

\begin{figure}
	\centering
	\includegraphics[width=9cm]{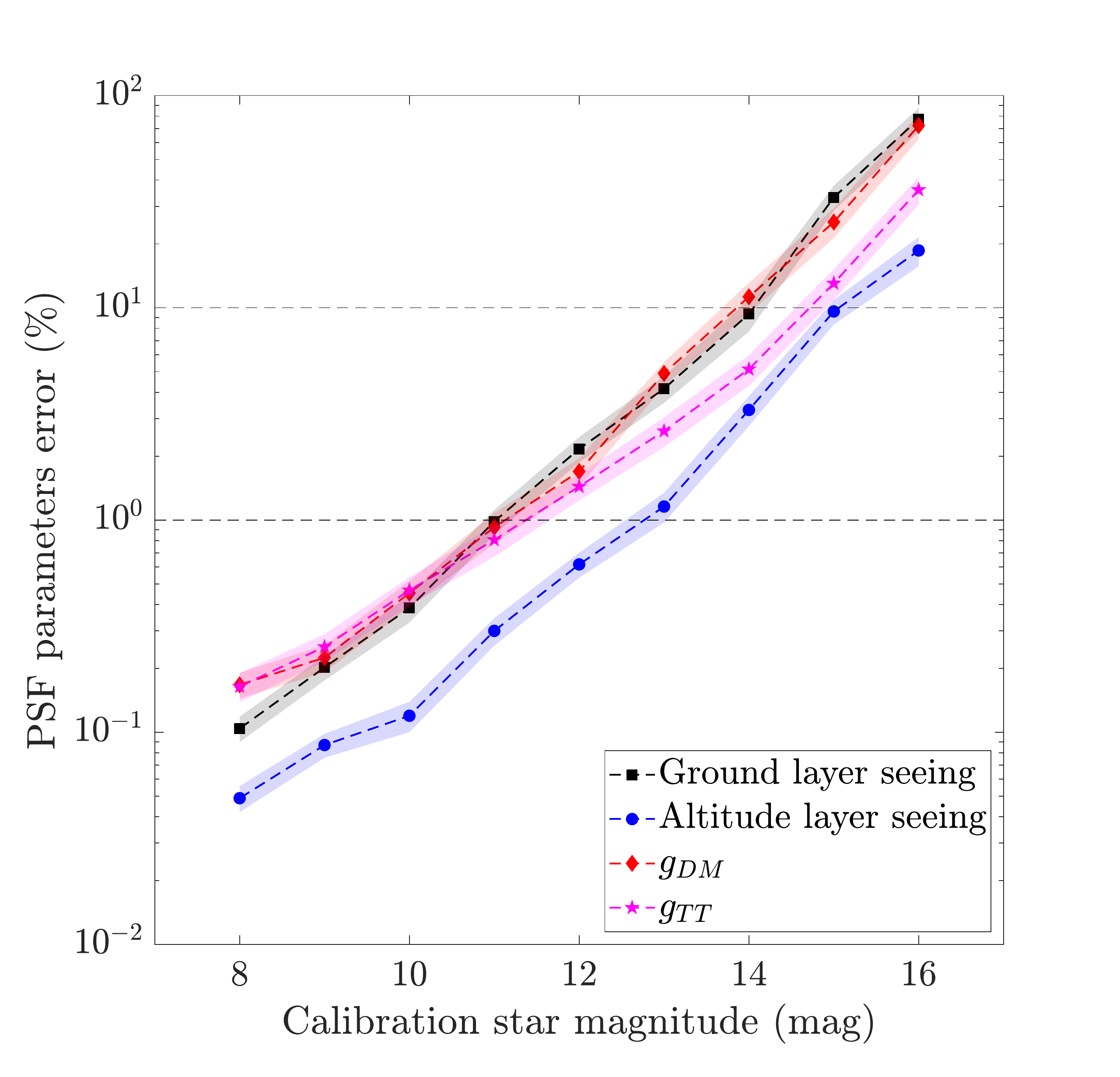}
	\caption{Accuracy on the PSF parameters estimation delivered by PRIME with respect to the calibration star magnitude with 30s of exposure time. Envelopes give the $1-\sigma$ standard-deviation.}
	\label{F:paramVmag}
\end{figure}

We have verified that the algorithm converges towards the exact solution ($\varepsilon = \varepsilon_\text{min}$) when removing electronic and photon noise. Then, we have tuned the star magnitude and contaminated the simulated image with noise before calibrating the four parameters (layers strengths plus WFS and TT optical gains) in order to assess the algorithm robustness to noise.

\begin{figure}
	\centering
	\includegraphics[width=9cm]{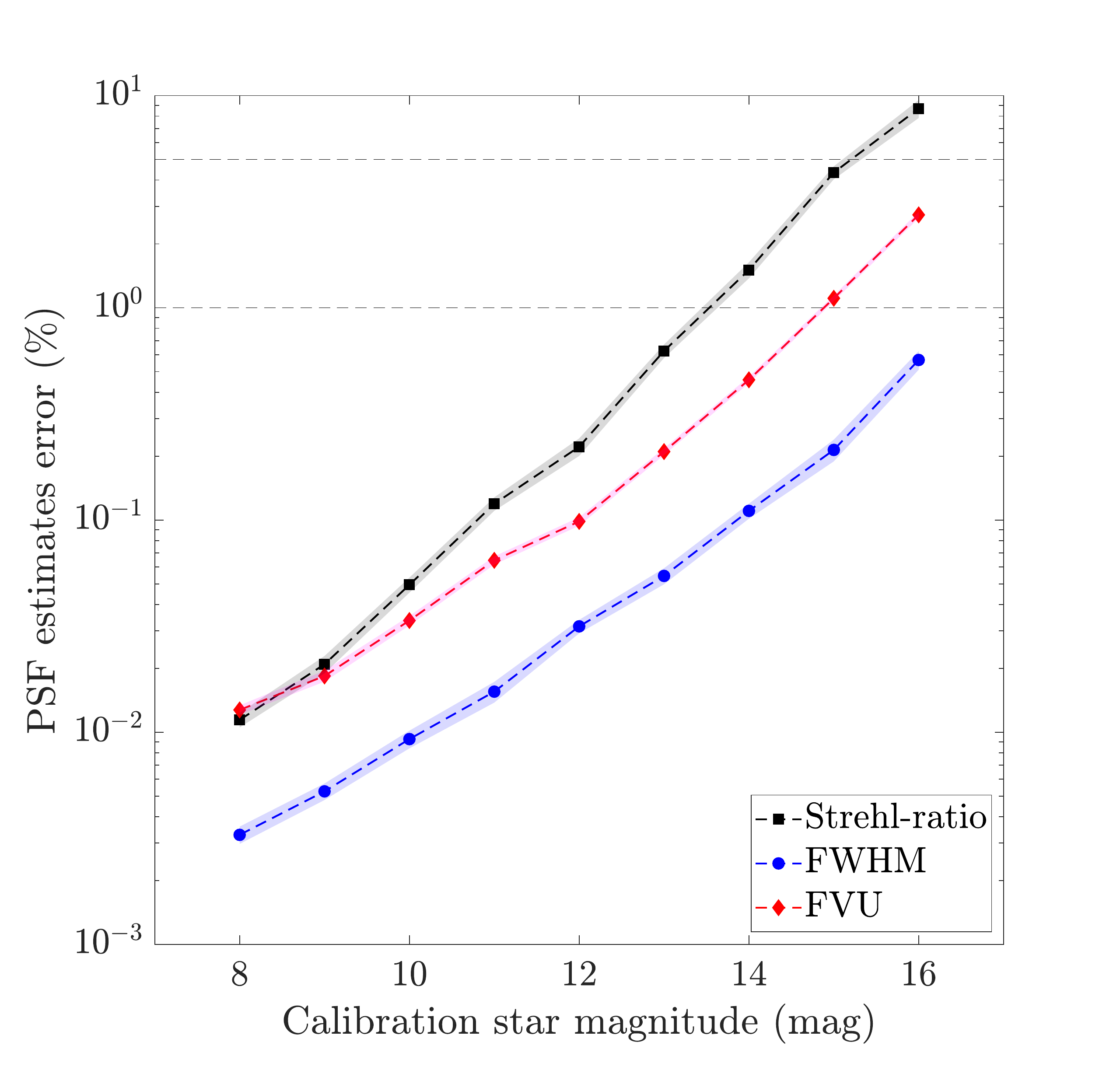}
	\caption{Accuracy on PSF estimates derived from the PRIME PSF best-fit with respect to the calibration star magnitude. Envelopes give the $1-\sigma$ standard-deviation.}
	\label{F:psfEstVmag}
\end{figure}

We report in Fig.~\ref{F:paramVmag} the accuracy on the adjusted parameters as a function of the calibration star magnitude. Fig.~\ref{F:psfEstVmag} reports the corresponding errors on PSF estimates, such as the SR, FWHM and Fraction of Variance Unexplained (FVU) that gives the overall PRIME PSF residual from the noise-free reference (\cite{BeltramoMartin2018_aniso}). SR is the most affected PSF estimate and reaches 1\% of error for a magnitude 13, for which we get 1\% on the altitude seeing estimation and 5\% on the other parameters.  Also, we learn that we can tolerate 5\% of error on optical gains to ensure a PSF modeling to better than 1\%-level of accuracy.\\

Finally, the FWHM estimation is more robust to noise compared to the SR evaluation. The main explanation of this is due to the PSF halo fitting that is strongly limited by the presence of noise, which affects the SR evaluation, compared to the brighter PSF core from which the FWHM is estimated. On top of that, the altitude seeing does impact the FWHM as well by introducing more or less anisoplanatism, which permits to enhance its estimation accuracy compared to the ground layer seeing that feeds the calculation of the PSF halo only. Although, this value modifies the PSF core morphology as well, this information is encoded into the telemetry and does not appear directly in the calculation of spatial filter $\text{K}_\text{DM}$ and $\text{K}_\text{TT}$. 

\section{PSF model calibration on NIRC2 at Keck II}
\label{S:NIRC2}

\subsection{PRIME NGS/LGS results}

In the following, we present PRIME results applied to the near infra-red imager NIRC2 at Keck II in narrow field mode with 9.94 mas/pixel sampling. We have processed 158 images collected simultaneously with the AO telemetry over three nights, in either NGS (August 1st 2013, March 14th 2017 and March 15th 2017, 80 images) (\cite{Wizinowich2000}) or LGS (March 14th 2017, 78 images) (\cite{Wizinowich2006}) mode.  We have systematically retrieved the seeing, optical gains, focus and astigmatism terms by using PRIME.  In LGS mode, the observations were taken with the NGS on-axis, which creates only focal anisoplanatism. This latter was not strong enough to identify a full $\cnh$ profile but does impact the SR and has been used in parallel with the PSF halo to constrain the seeing value. The sky image was cropped to 168 pixels to limit the noise propagation through the fitting process, which corresponds to 4 times the AO correction band.

\begin{figure}
	\centering
	\includegraphics[height=9cm]{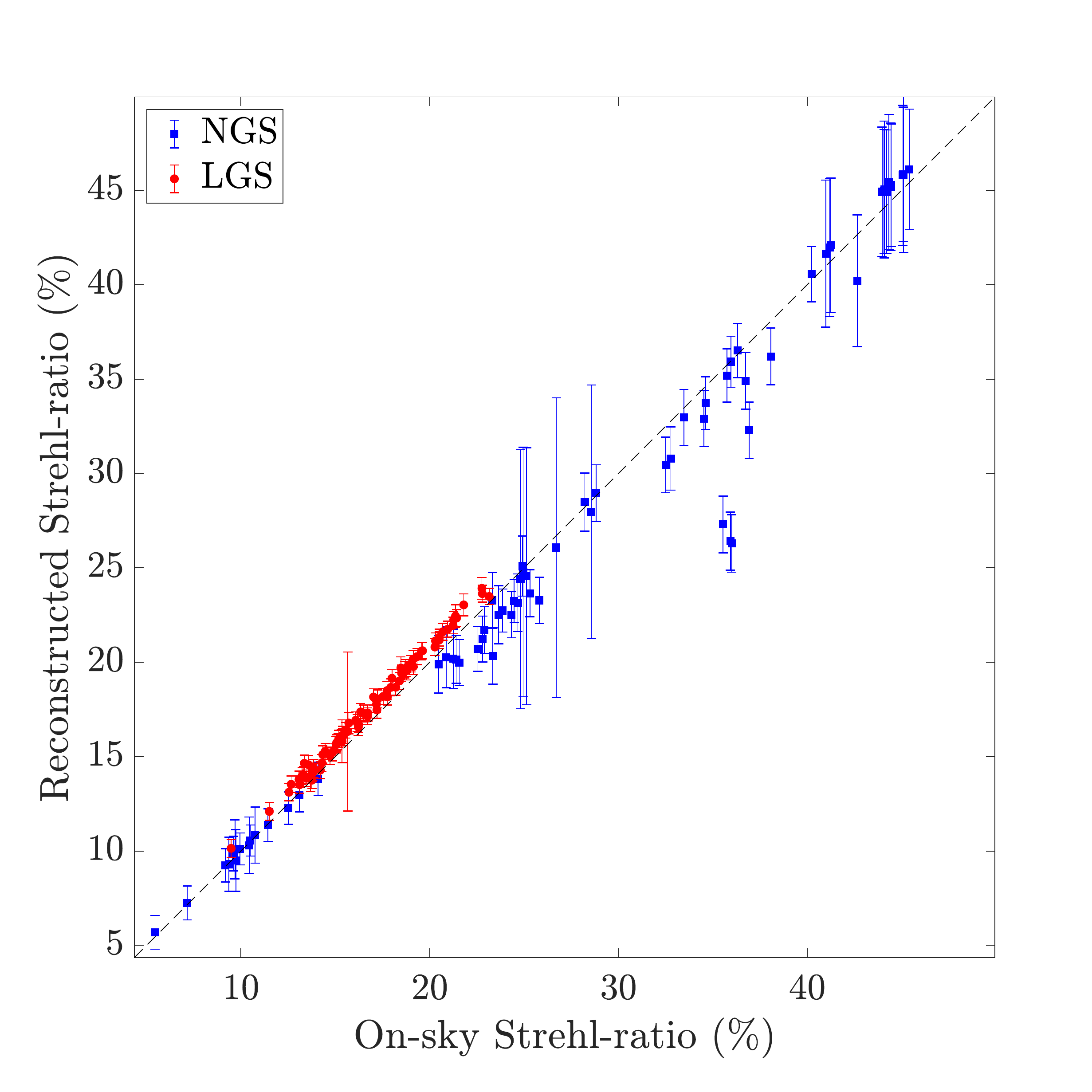}
	\includegraphics[height=9cm]{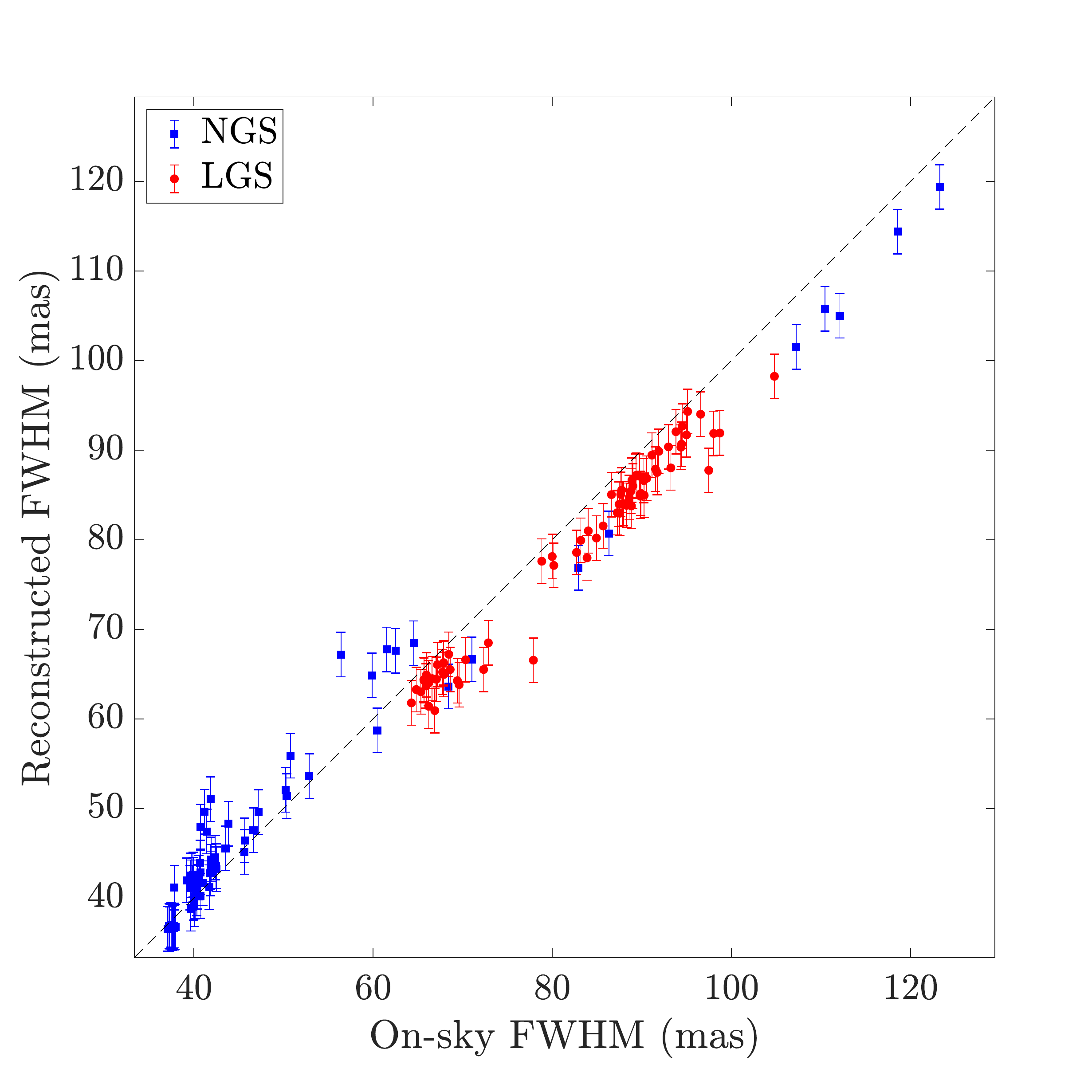}
	\caption{\textbf{Top~:} Reconstructed SR \textbf{Bottom~:} FWHM versus measurements over images obtained through classical PSF-R or PRIME in NGS mode. Values of Fe correspond to the WFS temporal frequency that only affects the PSF-R results only. Dashed line is the y=x line. }	
	\label{F:srSkyvsSRrec}
\end{figure}

\begin{figure}
	\centering
	\includegraphics[width=9cm]{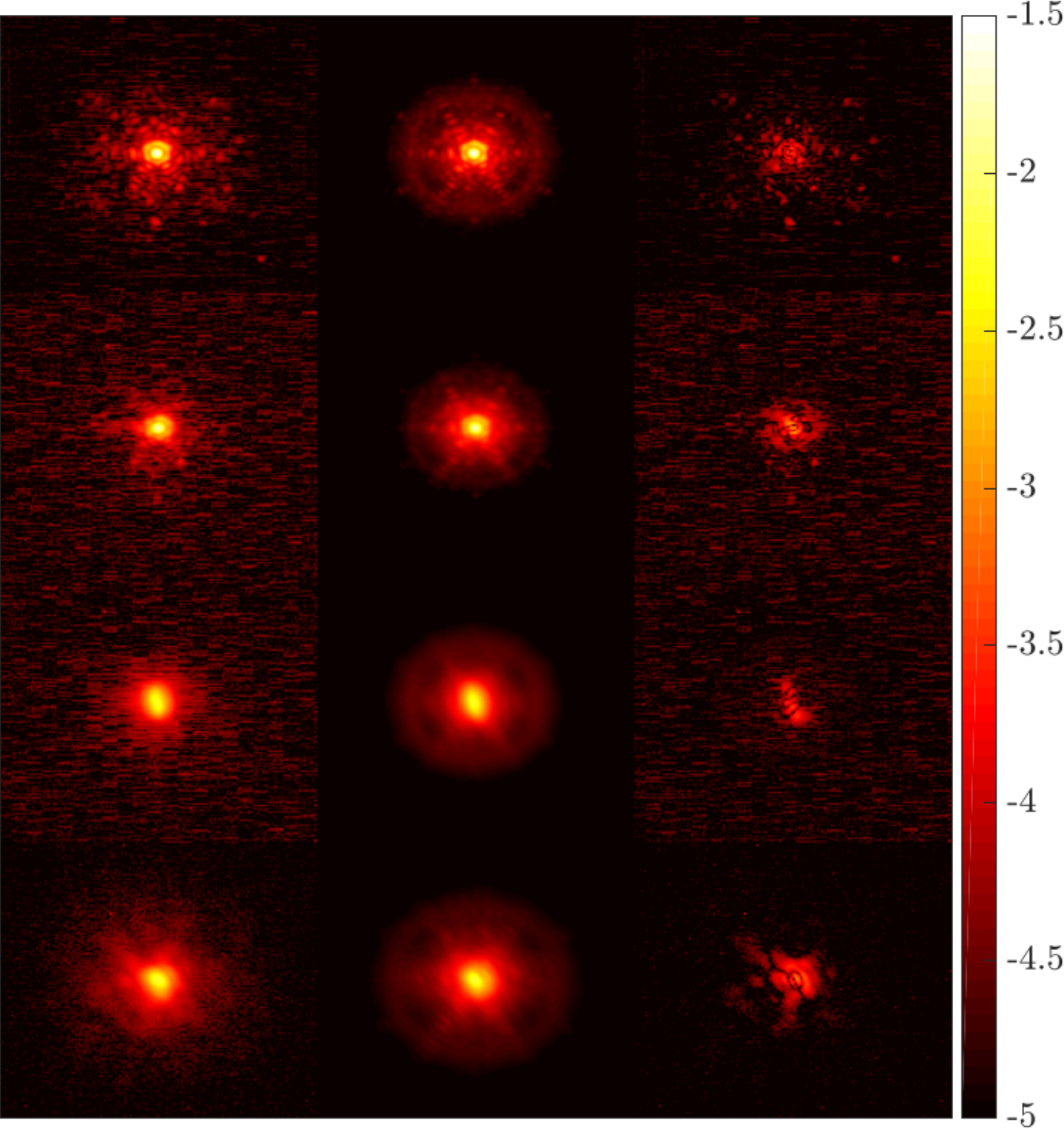}			
	\caption{Visual inspection of PRIME reconstruction results (left) on-sky PSFs (middle) PRIME best-fit (right) reconstruction residual. (1st row)  n0004  case in NGS mode in H-band (August 1st 2013) (2nd row) n0089 case, NGS mode, H-band (August 1st 2013) (3rd row) n0070 case, NGS mode, H-band (August 1st 2013) (4th row) n0086 case, LGS modes, K-band (March 14th 2017) }
	\label{F:psfr}
\end{figure}

We report in Fig.~\ref{F:srSkyvsSRrec} the reconstructed SR and FWHM versus measured values from images in NGS (1.65 $\mu$m) and LGS (2.12 $\mu$m) modes. PRIME reaches a vary good accuracy on SR and FWHM estimation of respectively 0.23\% $\pm$2.1\% and -0.75\,mas $\pm$ 3.5\,mas in NGS mode and -0.71\% $\pm$0.27\% and 3.2\,mas $\pm$ 1.9\,mas in LGS mode Furthermore, we provide in Fig.~\ref{F:psfr}  a visual inspection of sky and PRIME-reconstructed PSFs, that illustrates that on top of PSF metrics, the PSF morphology is well retrieved for various conditions of performance, even with strong PSF elongation due to wind-shake (case n0070).

\begin{figure*}
	\centering
	\includegraphics[width=8.5cm]{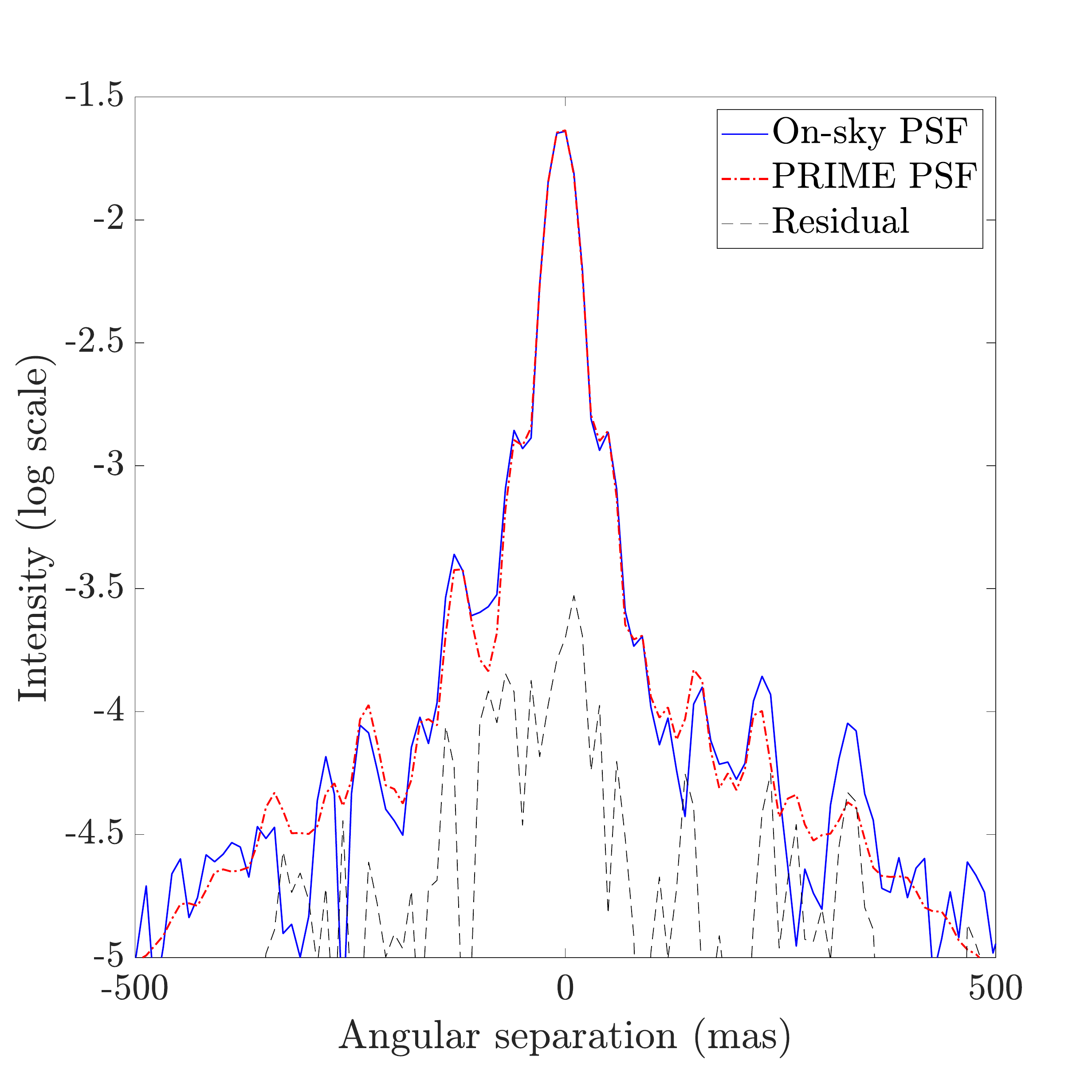}
	\includegraphics[width=8.5cm]{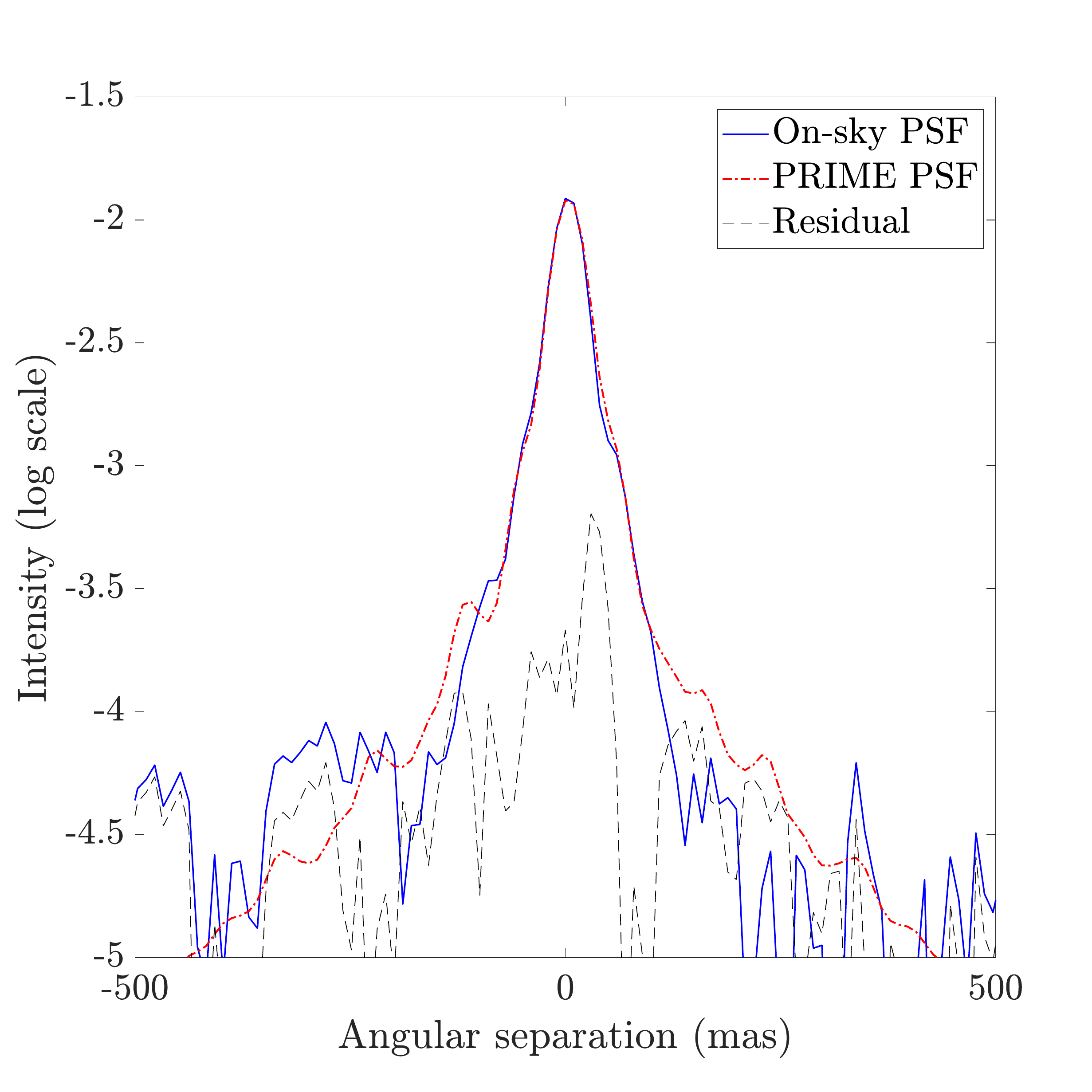}
	\includegraphics[width=8.5cm]{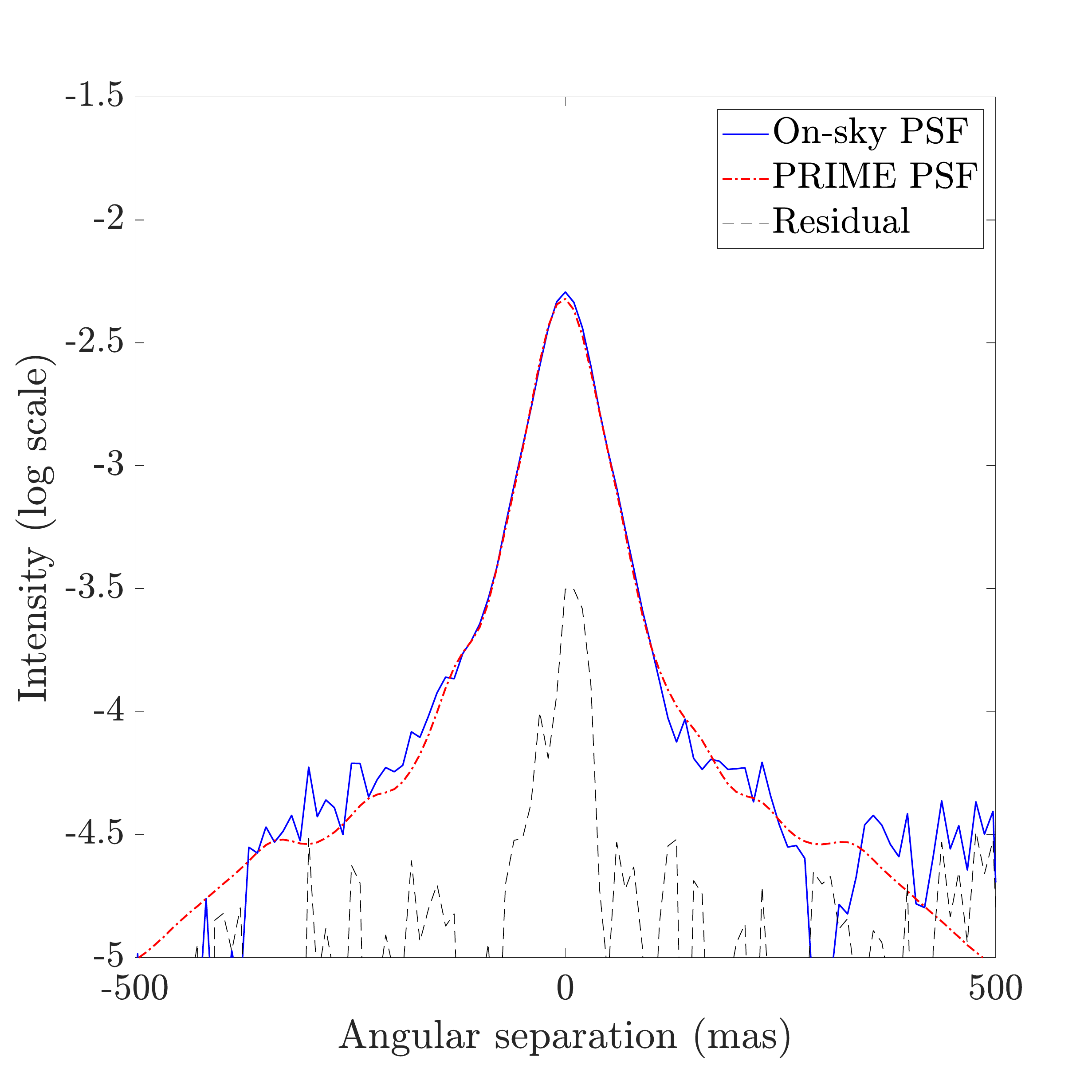} 
	\includegraphics[width=8.5cm]{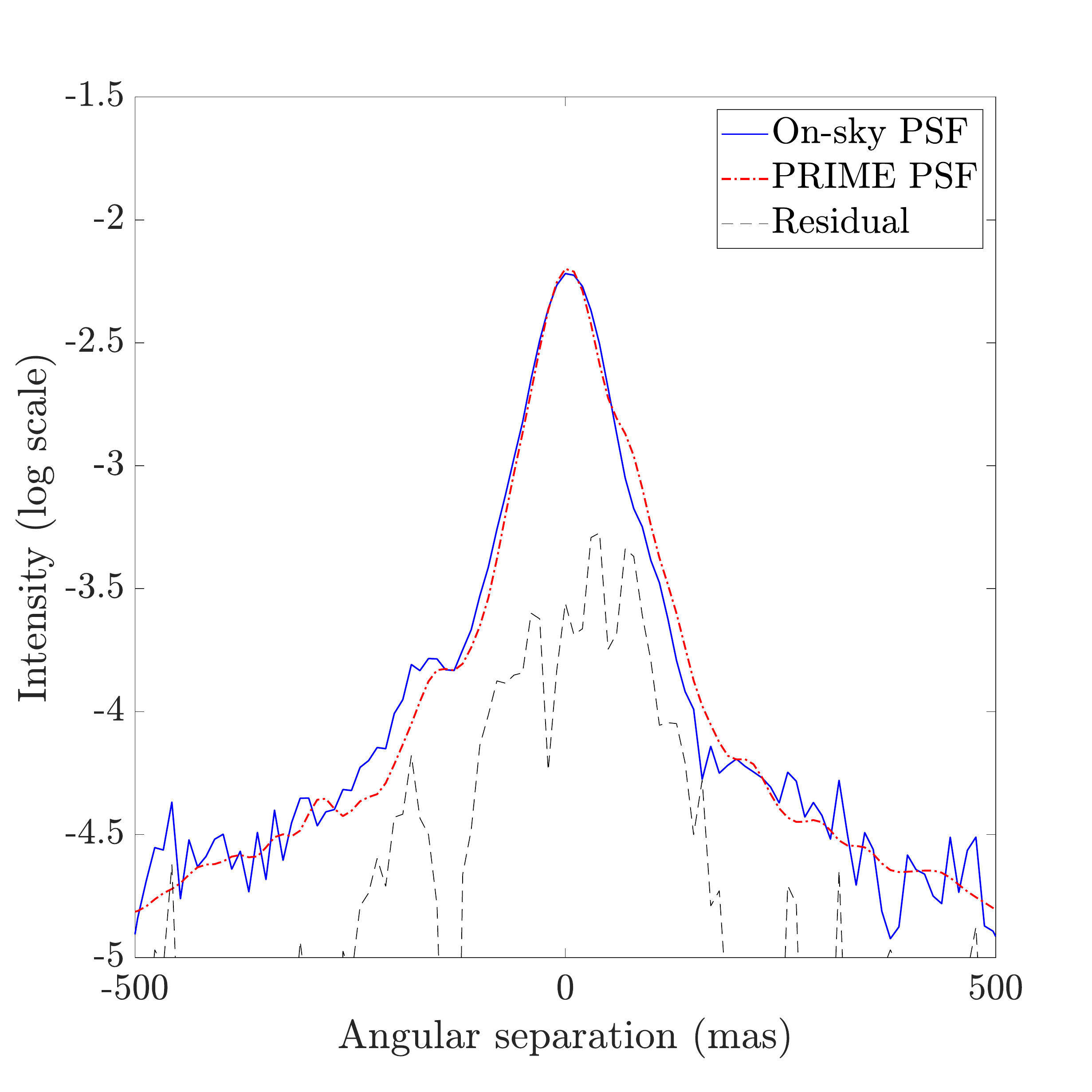} 
	\caption{Cuts of sky PSF compared to PRIME best-fit (\textbf{Top-left}) n0004  case in NGS mode in H-band (August 1st 2013)  (\textbf{Top-right}) n0089 case, NGS mode, H-band (August 1st 2013) (\textbf{Bottom-left}) n0070 case, NGS mode, H-band (August 1st 2013) (\textbf{Bottom-right}) n0086 case, LGS modes, K-band (March 14th 2017)  }
	\label{F:psfcut}
\end{figure*}


Such results were expected in the sense that we do best-fit the PSF model over the sky image whose morphology must be retrieved accurately. However, such an excellent agreement can only be obtained thanks to a representative and rich enough PSF model, which points out that classical PSF-R performance is only limited by the determination of key parameters. \\

\begin{figure}
\centering
	\includegraphics[height=9cm]{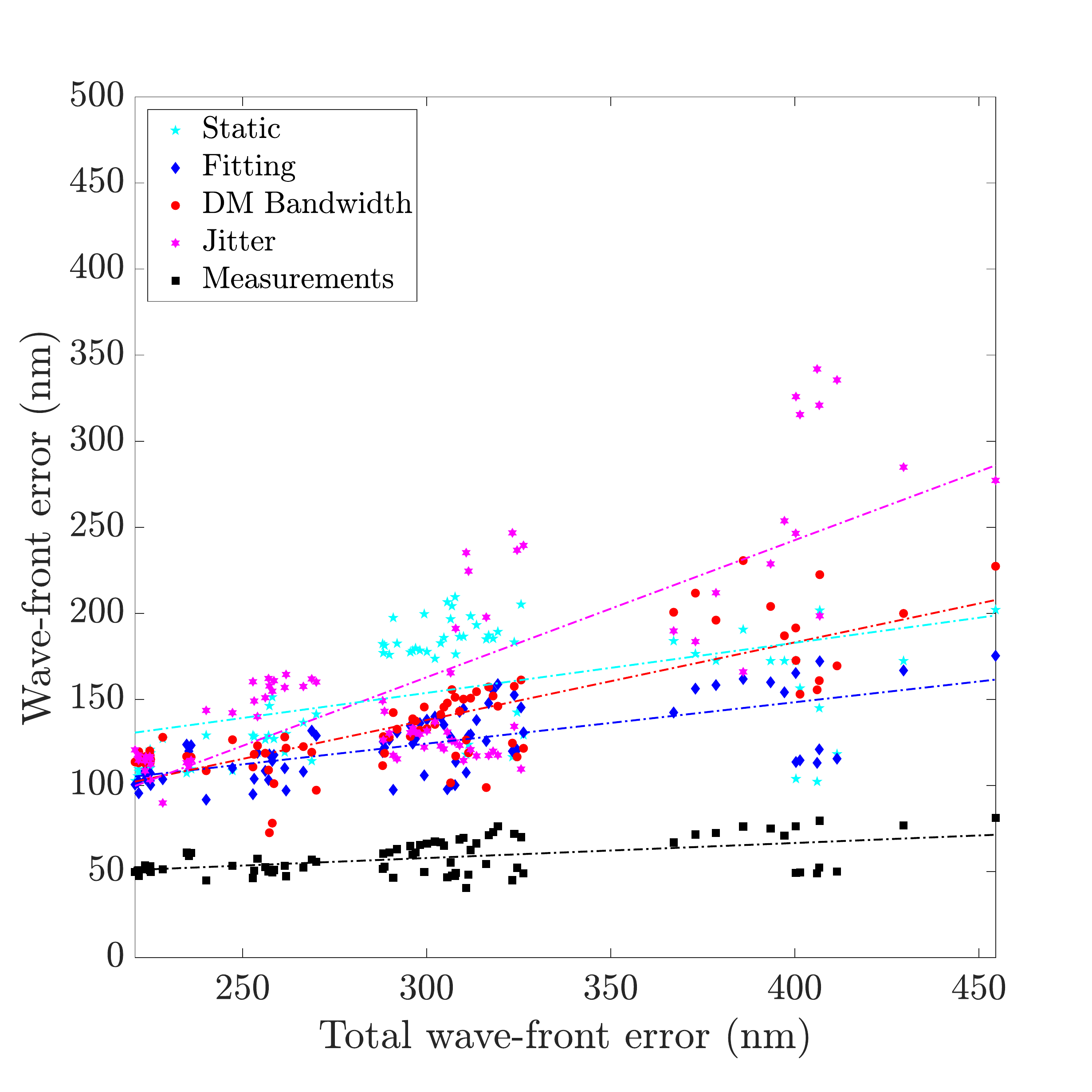}
	\includegraphics[height=9cm]{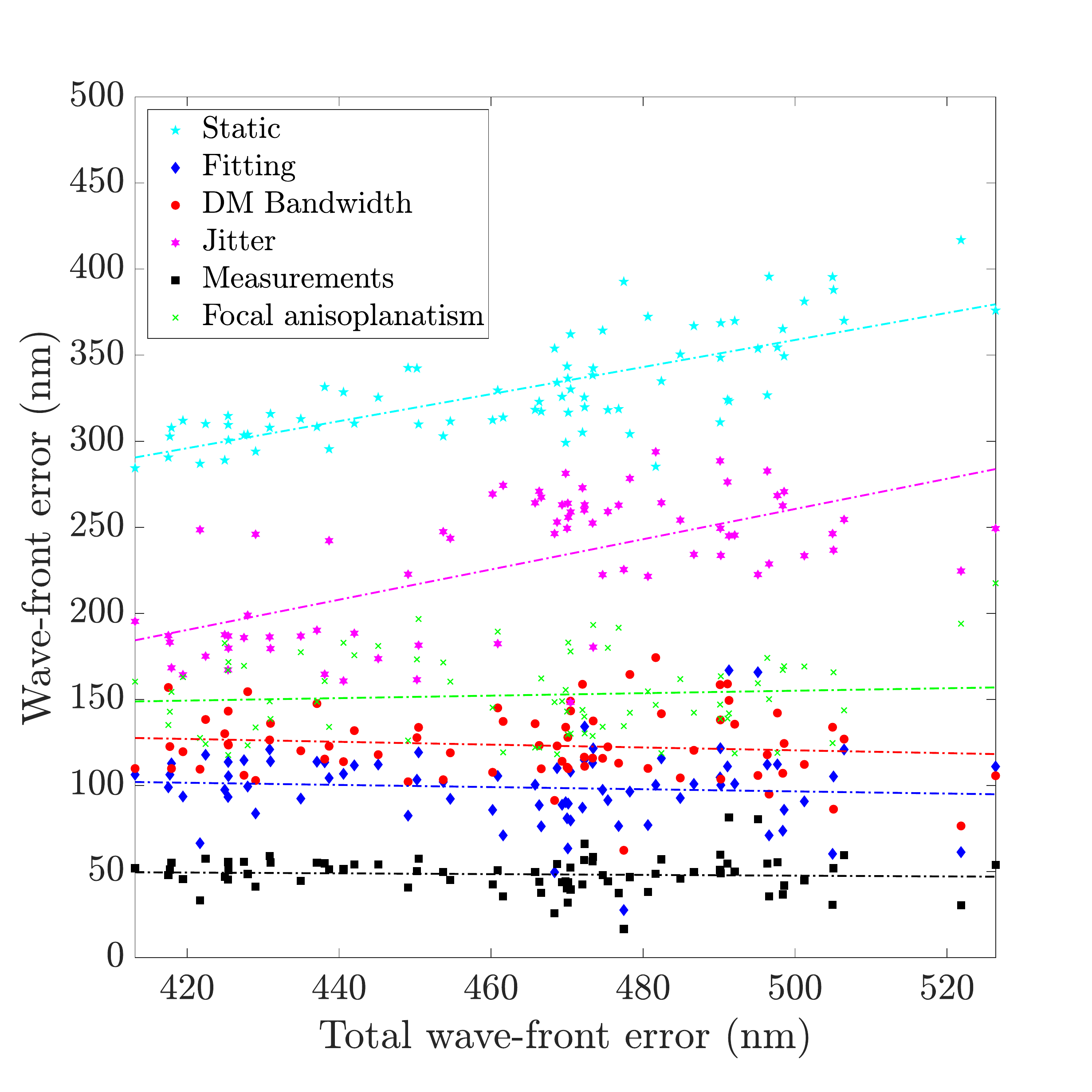}
	\caption{Detailed wavefront error breakdown with respect to the total error. Measurements error refers to WFS noise and aliasing. \textbf{Top~:} NGS \textbf{Middle~:} LGS results.}
	\label{F:wfe}
\end{figure}

\begin{figure}
\centering	
	\includegraphics[height=9cm]{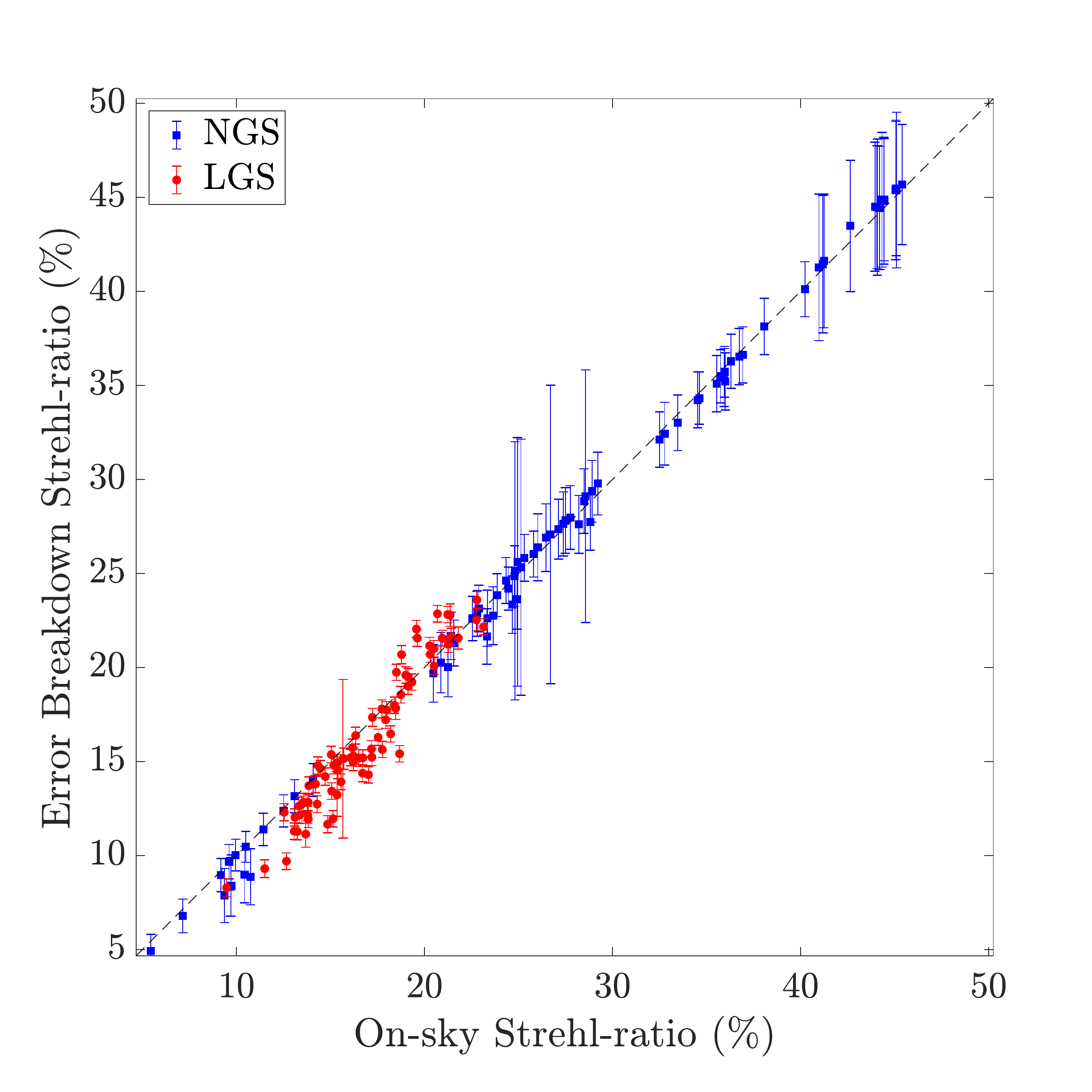}
	\caption{Maréchal SR derived from the error breakdown versus the sky SR.}
	\label{F:srMarvSRsky}
\end{figure}

We have used equation~\ref{E:wfe} to derive the wavefront-error breakdown based on PRIME outputs. We report the decomposition of the total wavefront error in Fig.~\ref{F:wfe} in NGS and LGS modes with respect to the AO performance. Absolute values are verified by computing the Maréchal SR from the total variance and compare it to the value estimated from the image. Results presented in in Fig.~\ref{F:srMarvSRsky} shows an excellent agreement and provide confidence into the error breakdown estimation.

Poor AO performance are explained by strong jitter effects that dominates the turbulence in terms of power and therefore residual error. On top of that wind-shake effects participate to boost the PSF residual jitter. To overcome this limitation as much as possible, Keck I AO system is equipped with TRICK, a Near infra-red tip-tilt sensor that is designed to improve the tip-tilt estimation (\cite{Wizinowich2016}). 

LGS observations also suffered by very large low-order aberrations , i.e. focus and astigmatism, that explains the poor performance contrary to NGS cases for which static terms were dominated by NCPA residual. The Low-Bandwidth Wavefront Sensor (LBWFS) (\cite{Wizinowich2006}) is supposed to capture these low-order modes and update the AO loop at a low temporal frame rate to correct for them. However, the LBWFS telemetry is not yet available into the Keck Telemetry Recording System (TRS) and additional investigation must be led to verify if including the LBWFS measurements into Eq.~\ref{E:otfDM} could enhance the PSF reconstruction. Nonetheless, if these modes are effectively perceived by the LBWFS, they must be compensated and not affect the PSF as strongly as we see on sky images, indicating that they are created by another effect, as potentially variations of static aberrations. We also notice that these static errors grow up while the atmospheric fitting error and the focal anisoplanatism decreases, i.e. for lower values of seeing. This observation indicates that static errors can be potentially introduced by variations of the WFS optical gain, which modify WFS reference slopes and may degrade the compensation of NCPA. \\

We observe a similar anti-correlation between the residual jitter and the seeing. When the wind blows faster, it shakes the telescope and the dome, which impacts the PSF as we saw in Fig.~\ref{F:psfr} with the case n0070.  In LGS mode, the DM bandwidth error grows up despite the diminution of the seeing value, which pinpoints that this augmentation is due to a greater wind speed value. Observations lead to the conclusions that the good seeing conditions can be reached in presence of fast wind, but such good seeing conditions can not be optimally exploited because the telescope wind-shake.\\

\subsection{PSF prediction}
\label{SS:prediction}
We have compared PRIME retrieved parameters to reference values to understand in such way we can calibrate the PSF model regarding the observing conditions when no point source is available in the field. We have focused on all the NGS data we have and a sub-sample of 40 LGS cases acquired on the 14th March 2017, in order to learn about the parameters evolution and predict the PSF from the telemetry on a test sample of the 38 remaining LGS data. All the errors bars in the following are obtained from the Levenberg-Marquardt best-fitting procedure that deliver the estimates precision from the PSF model gradients calculated empirically.

\begin{figure}
\centering
	\includegraphics[height=9cm]{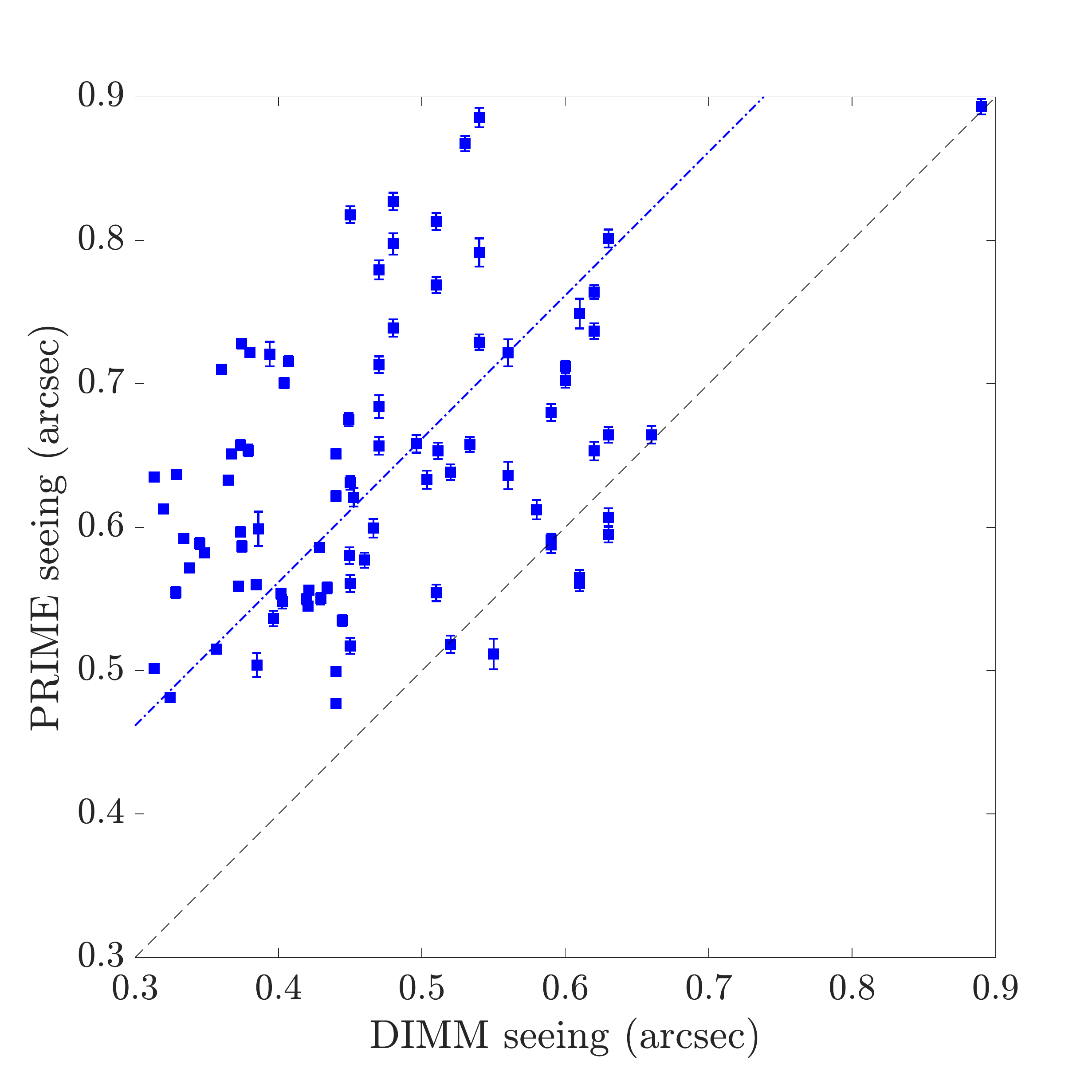}
	\caption{Seeing estimates from PRIME with regard to DIMM measurements. Dash line corresponds to y=x and the dotted line is the linear regression showing a bias of 0.06 arcsec.}
	\label{F:seeingPRIMEvsDIMM}
\end{figure}

We report in Fig.~\ref{F:seeingPRIMEvsDIMM} the seeing retrieved with PRIME on both NGS and LGS images compared to the DIMM measurements at MaunaKea. It highlights a bias of 0.06" and a large discrepancy of 0.14" rms. We can consequently compensated for this bias in the PSF model to improve the PSF halo reconstruction but the large discrepancy will eventually limit the PSF prediction. Alternative seeing estimation must be therefore considered from the AO telemetry (\cite{Jolissaint2018,Andrade2018}). Our next step will be to combine seeing estimation method from AO control loop data (\cite{Jolissaint2018,Andrade2018}) and PRIME to constrain more efficiently the seeing determination and enhance the robustness to noise.

\begin{figure}
\centering
	\includegraphics[height=9cm]{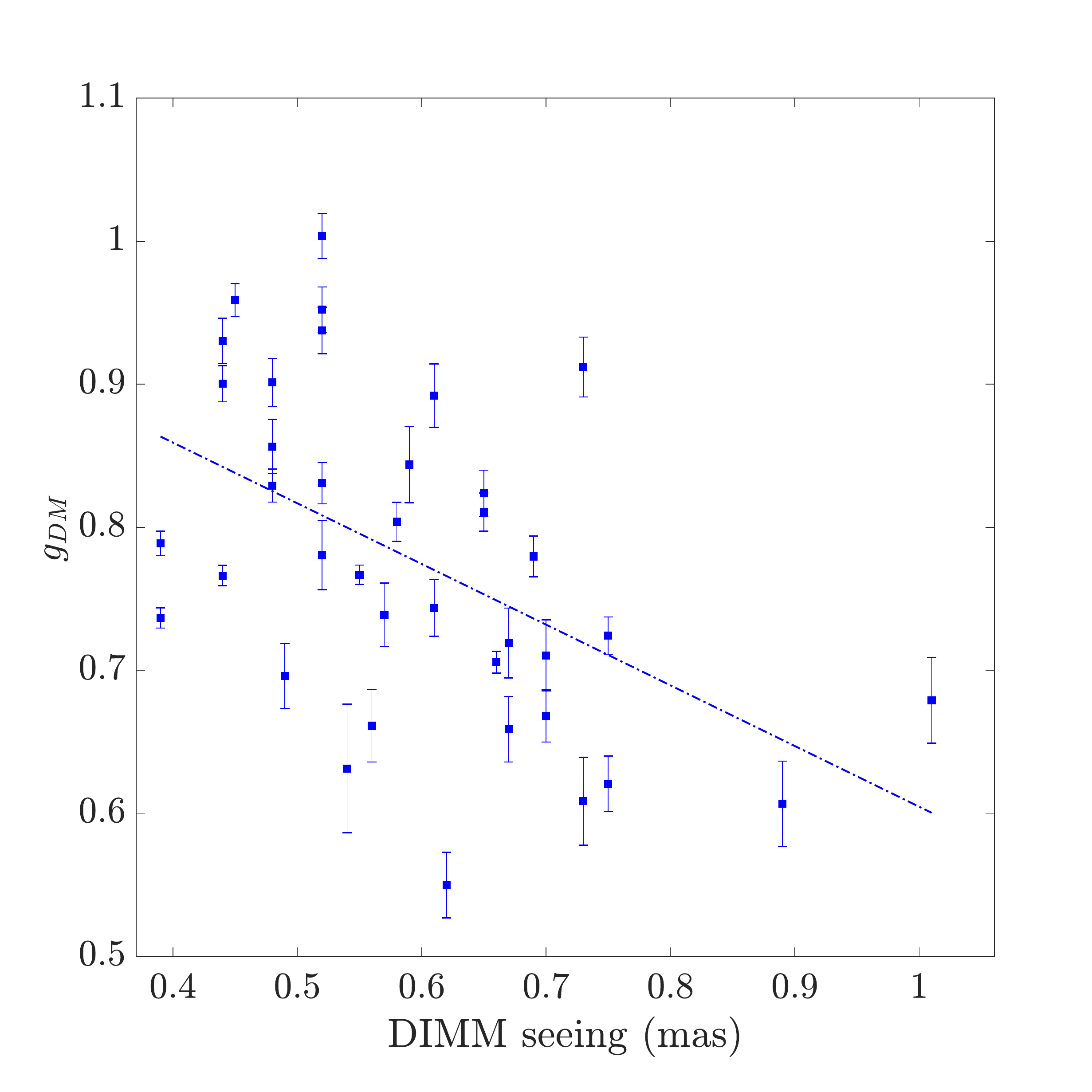}
	\caption{Optical gain estimates with PRIME versus the DIMM seeing over the 40 LGS learning sets.}
	\label{F:gDMVseeingDIMM_LGS}
\end{figure}

We report in Fig.~\ref{F:gDMVseeingDIMM_LGS} the evolution of the retrieved optical gain versus the DIMM seeing. We observe a decreasing according to the following empirical law $g_\text{DM} = 1.03 - 0.4243\times \text{s}_\text{DIMM}$, where $\text{s}_\text{DIMM}$ is the DIMM seeing. We observe a slighter decreasing for the TT optical gain for which we have $g_\text{TT} = 0.98 - 0.037\times \text{s}_\text{DIMM}$. Once again, the samples cloud is quite dispersed, but the trend is clear enough to be calibrated and compensate for the reconstruction bias.

\begin{figure}
\centering
	\includegraphics[height=9cm]{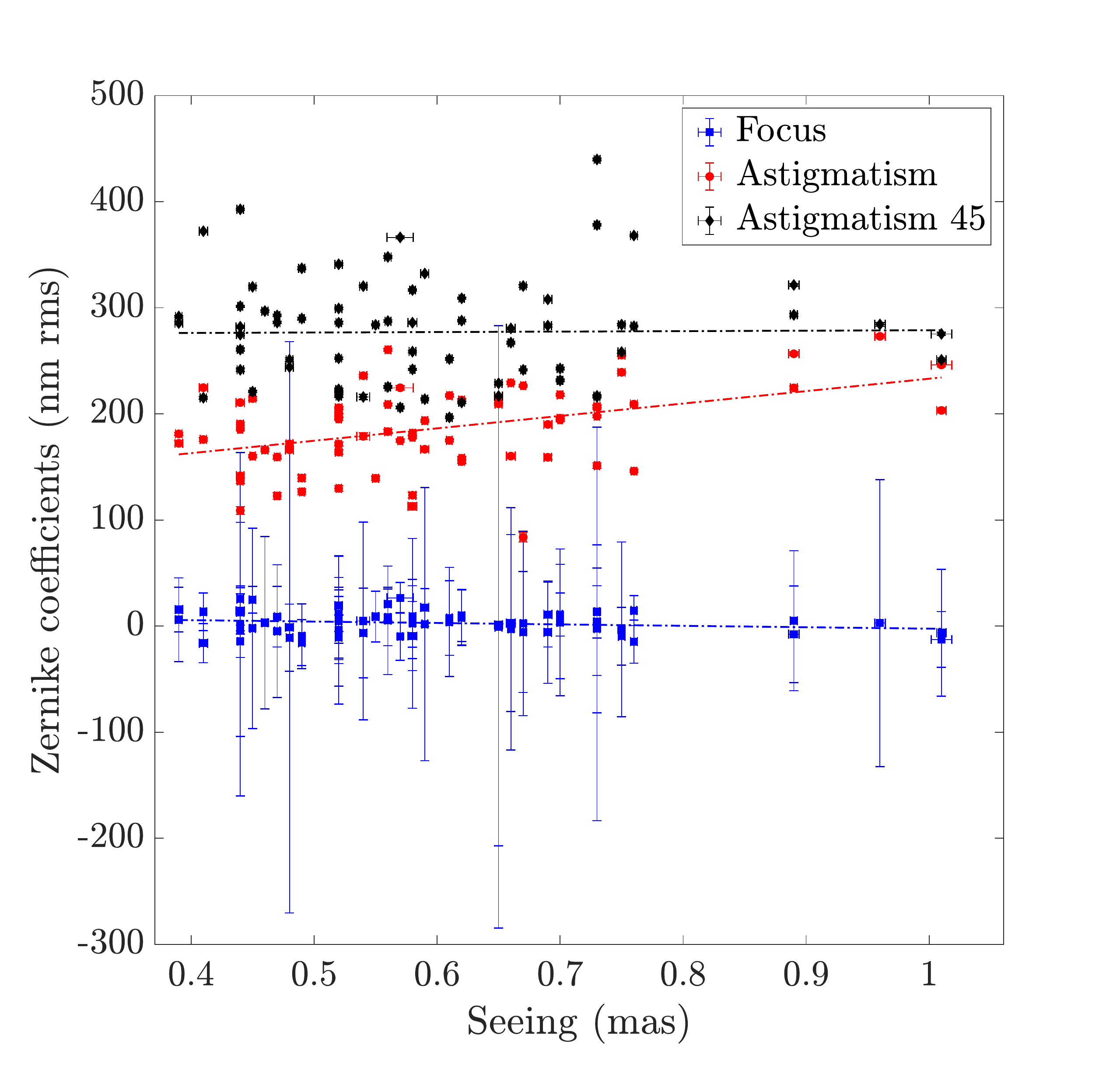}
	\caption{Static terms retrieved with PRIME versus the DIMM seeing over the 40 LGS learning sets.}
	\label{F:coefsVgDM_LGS}
\end{figure}

Finally, we represent in Fig.~\ref{F:coefsVgDM_LGS} the three Zernike coefficients estimated by PRIME with respect to the DIMM seeing whose the dependency is modeled linearly with $a_4 = -4.7 + 1.68\times \text{s}_\text{DIMM}$, $a_5 = 48 + 110\times \text{s}_\text{DIMM}$ and $a_6 = 221 + 29.9\times \text{s}_\text{DIMM}$

\begin{figure}
\centering
	\includegraphics[height=9cm]{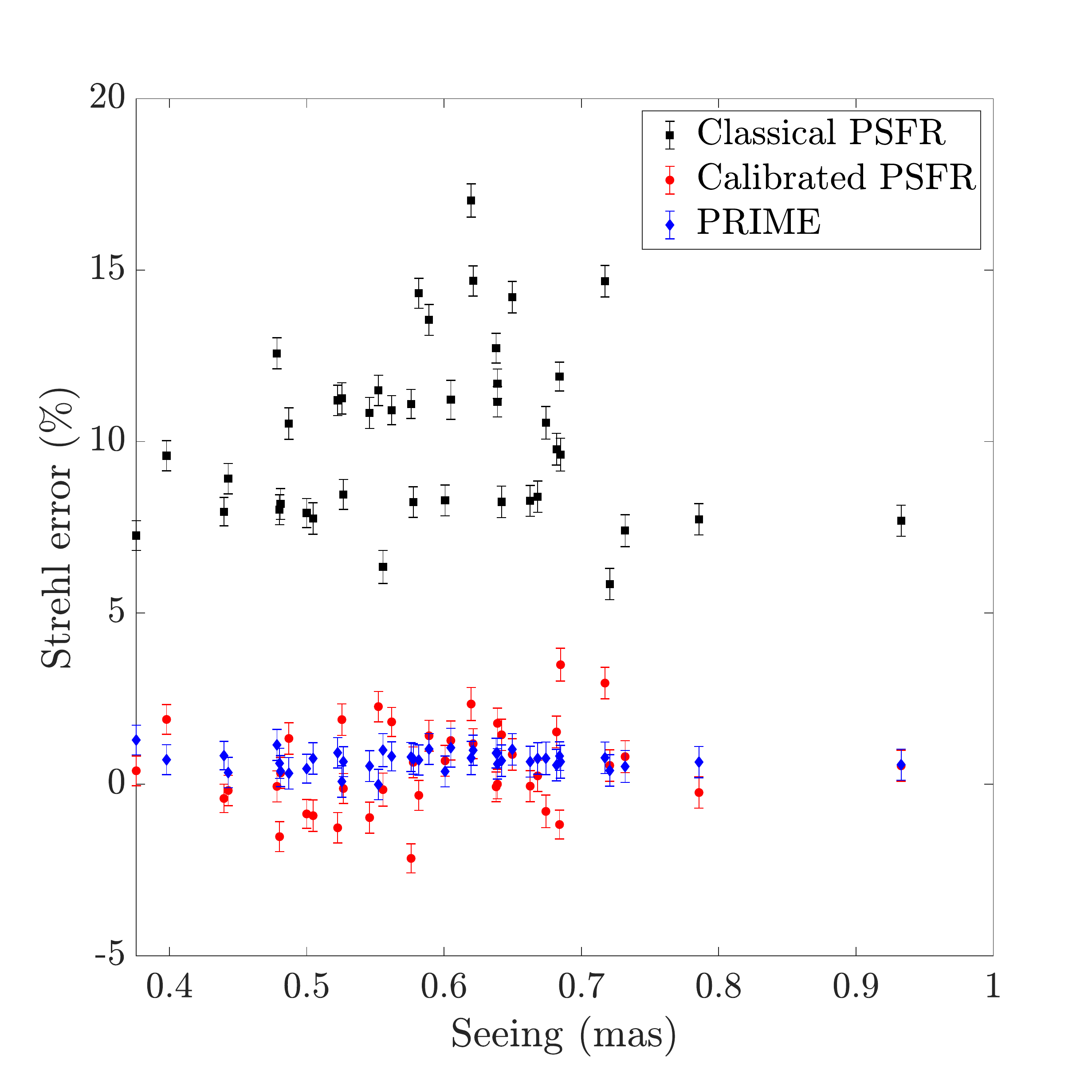}
	\includegraphics[height=9cm]{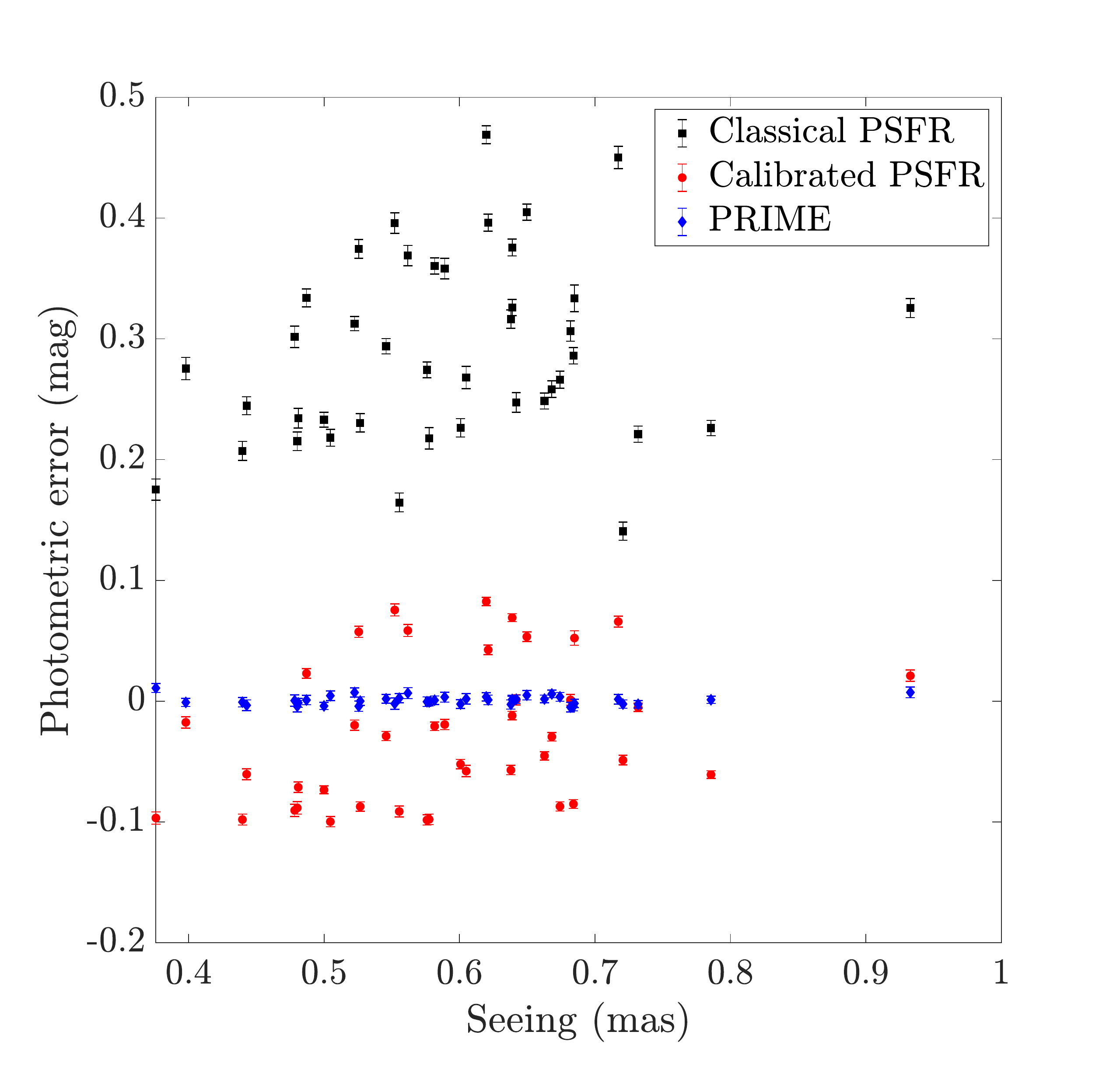}
	\caption{\textbf{Top} Strehl ratio error \textbf{Bottom} photometric error versus the seeing for the 40 LGS test samples. The figure compares the classical PSF-R (optical gains set to one and use of the DIMM seeing), calibrated PSF-R (use of predictive parameters evolution laws) or PRIME (focal-plane best-fit).}
	\label{F:magErrorVseeing}
\end{figure}

Furthermore, we have compared PRIME to AO telemetry-based only PSF-R on the LGS test case with two configurations: either we neglected system parameters variations with respect to the seeing (g$_\text{DM}$ = g$_\text{DM}$ = 1, a$_4$=a$_5$=a$_6$ = 0), or we use the empirical laws deduced on the learning set. We report in Fig.~\ref{F:magErrorVseeing} the SR reconstruction error and the photometric accuracy with respect to the seeing. Photometry is obtained from the residual of the scaling of the reconstructed PSF over the sky image. We report in Tab. \ref{T:statsSR} the corresponding error statistics in terms of bias and standard-deviation. FWHM results are very similar between the three methods, but slightly better with PRIME and the calibrated PSF-R compared to the classical PSF-R, advocating that the PSF FWHM is already well estimated from the AO telemetry only.

Results highlight that the parameters calibration allows to correct for the bias on estimates with a similar level compared to PRIME and decreases the discrepancy by 50\% compared to the uncalibrated PSF-R. If point source are available in the field to calibrate the PSF model using PRIME, we can expect a photometric accuracy at the level of 0.004 mag, that degrades to 0.06 when relying on parameters calibration only with the present status of the system and the PRIME method. Such results are already promising to deploy PRIME for constraining the PSF model in the image post-processing pipeline as we plan to do. On top of that, by deploying PRIME or more and more archive data, we will be able to improve our understanding of the AO system and enhance its performance, make it more stable and more predictable, which will also make the system parameters prediction more robust and reliable.

\begin{table}
	\centering
	\caption{Statistics on SR and photometric error reconstruction using either the classical PSF-R (optical gains set to one and use of the DIMM seeing), calibrated PSF-R (use of predictive parameters evolution laws) or PRIME (focal-plane best-fit) over the LGS test images.}	
	\begin{tabular}{|c||c|c||c|c|}
		\hline
		& \multicolumn{2}{c||}{$\Delta$ SR (\%)} & \multicolumn{2}{c|}{$\Delta$ mag (mag)} \\
		\hline
		& Mean & RMS & Mean & RMS\\
		\hline
		Classical PSF-R  & 8  & 3.2 & 0.22  & 0.12  \\
		\hline
		Calibrated PSF-R &  -0.8 & 2.0  & -0.03 & 0.06\\
		\hline
		PRIME 			 & 0.7 & 0.3 & 0.0009 & 0.004 \\
		\hline
	\end{tabular}
	\label{T:statsSR}
\end{table}

\section{Application of PRIME to tight-binaries}
\label{S:binaries}

\subsection{Impact of noise on stellar parameters estimation}
\label{SS:binNoise}

\begin{figure*}
	\centering
	\includegraphics[width=10cm]{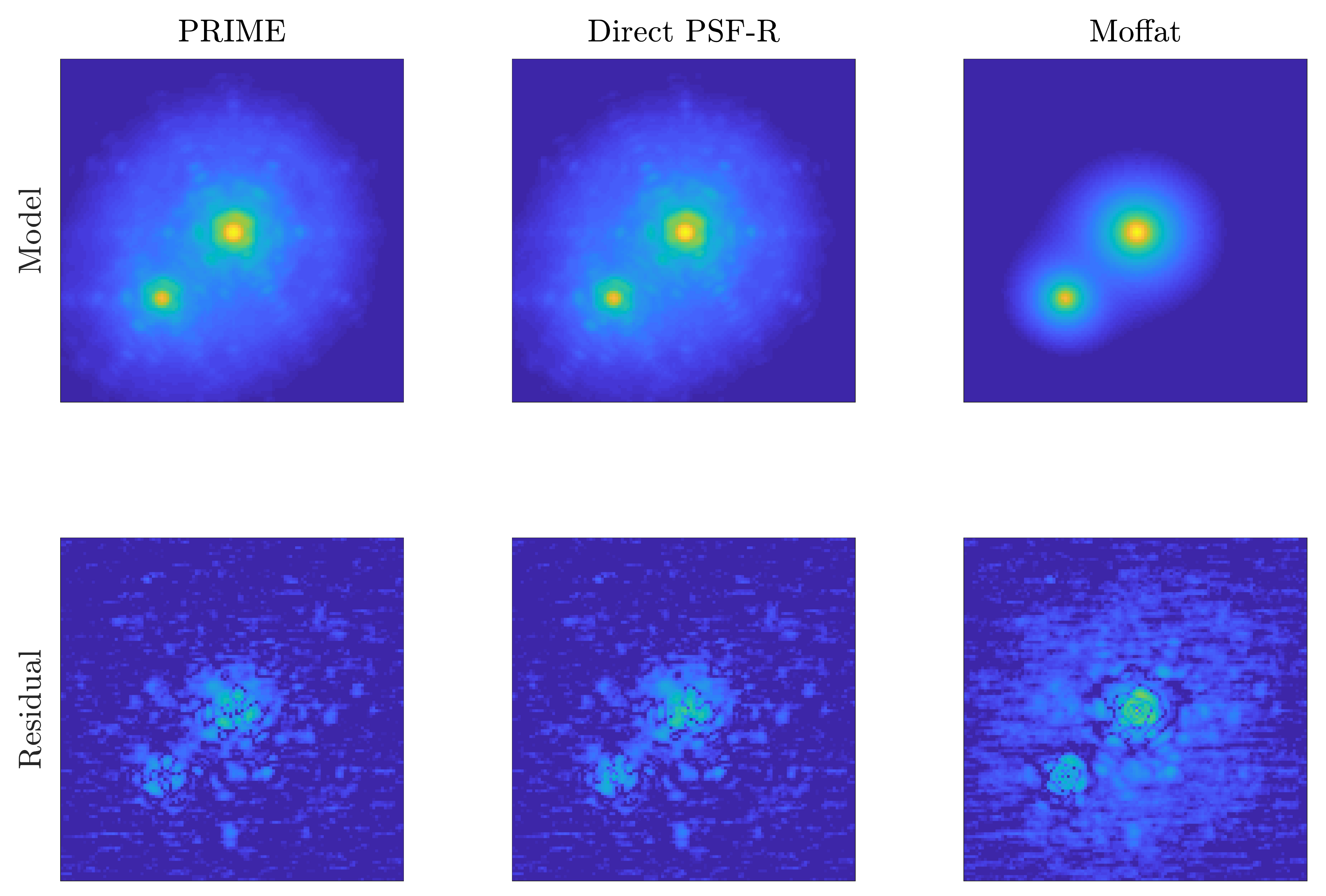}
	\includegraphics[width=7.5cm]{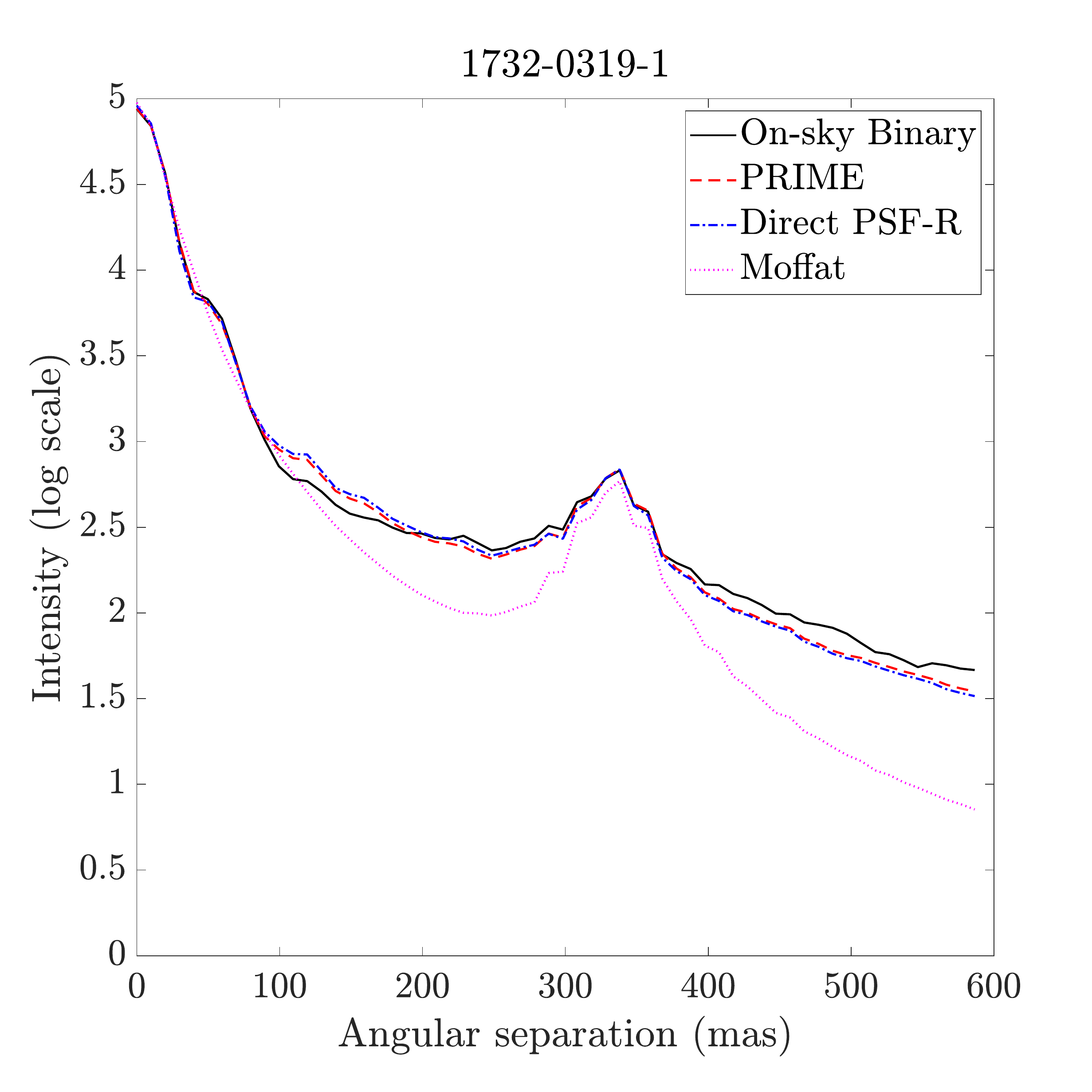}
	\caption{1732-03910-1 case. \textbf{Left panel:} Top images are respectively PRIME and Moffat fitting; bottom figures give the fitting residual over sky images. \textbf{Right panel:} Azimuthal average of the binary image averaged over the five acquisition with overplotting of best-fit obtained with PRIME, PSF-R and a PSF Moffat model.}
	\label{F:1732}
\end{figure*}

\begin{figure*}
	\centering
	\includegraphics[width=10cm]{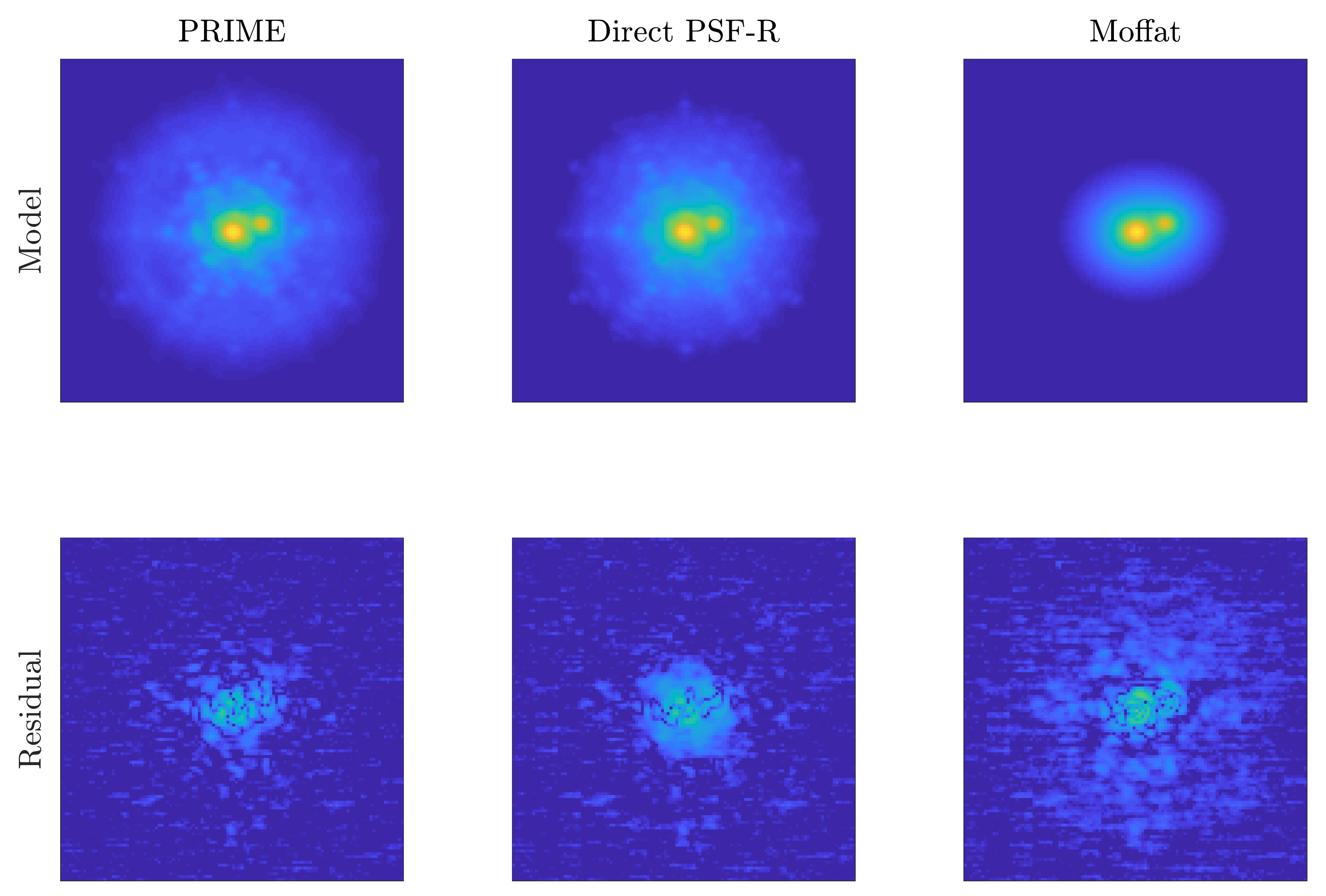}
	\includegraphics[width=7.5cm]{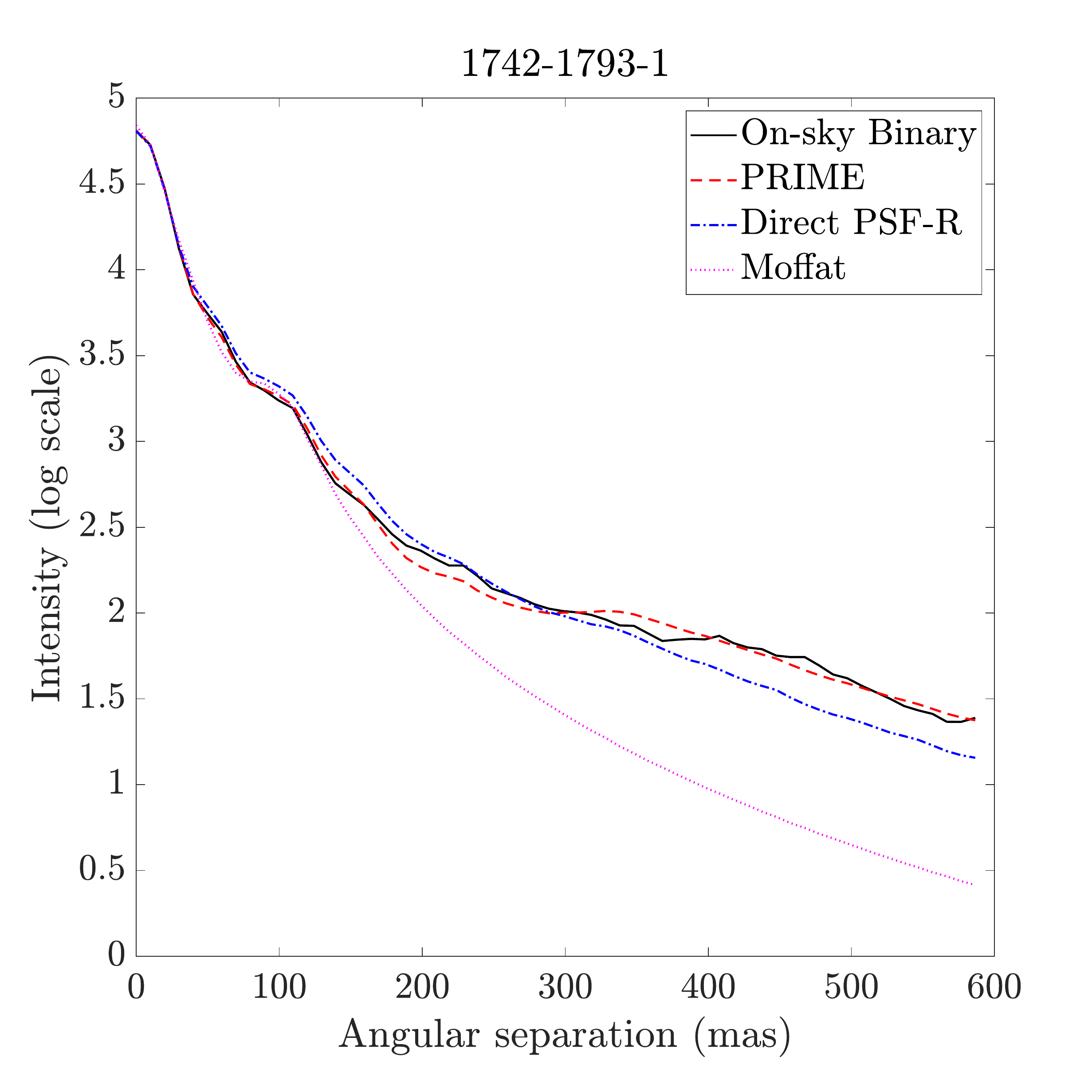}
	\caption{1742-1793-1 case. \textbf{Left panel:} Top images are respectively PRIME and Moffat fitting; bottom figures give the fitting residual over sky images. \textbf{Right panel:} Azimuthal average of the mean binary image over the five acquisition with overplotting of best-fit obtained with PRIME, PSF-R and a PSF Moffat model.}
	\label{F:1742}
\end{figure*}

We handle in this section images of binary data 1732-0319-1 and 1742-1793-1 acquired with NIRC2 in narrow field band by using the FeII filter (1.6455 $\mu$m), with a good Signal-to-noise ratio (S/N) over 5 frames of 50s exposure each. We have measured the stability of the photometry and astrometry measurements by using either PRIME, or predicted PSF-R or a Moffat model. We present in Figs.~\ref{F:1732} and~\ref{F:1742} a 2D visual comparison of best-fitted models as well as azimuthal profile averaged out over the five frames for the 1732-03910-1 and 1742-1793-1 cases. Present results confirm that PRIME can estimate jointly PSF and stellar parameters efficiently. A single Moffat function is not accurate enough to reproduce the entire PSF shape and the PSF-R direct model inputs were not sufficiently well identified to reach the same residual level than PRIME. We particularly notice that PRIME has played on the seeing value to adjust the PSF wings level. \\

\begin{table}
	\centering
	\caption{Statistics on estimation stability of the binary separation and differential photometry regarding the PSF model. Values are averaged over the two objects.}	
	\begin{tabular}{|c|c|c|c|}
		\hline
		& FVU (\%) & $\sqrt{\Delta \alpha_1^2 + \Delta \alpha_2^2}$ ($\mu$as) & $\sqrt{\Delta p_1^2 + \Delta p_2^2}$ (mag)  \\		
		\hline
		Moffat & 1.67$\pm$0.34 	& 473 $\pm$ 18  	& 0.014 $\pm$ 0.001 \\
		\hline
		PSF-R  & 0.77$\pm$0.16 	& 333 $\pm$ 63   	& 0.001 $\pm$ 0.001 \\
		\hline
		PRIME & 0.49$\pm$0.05	& 263 $\pm$ 31  	& 0.008 $\pm$ 0.0005  \\	
		\hline
	\end{tabular}
	\label{T:stats2013}
\end{table}

Thanks to PRIME, we gain by having more degree of freedom to achieve better stellar characterization, but we do potentially increase the noise propagation though the minimization loop. Because the S/N was very good on the present images, we can artificially play on the object digital intensity to change its apparent magnitude regarding NIRC2 photometric specifications and simulate the observation we would have with a different S/N. The goal of our approach is to capture how the metrics stability degrades with respect to the equivalent magnitude and the PSF model. 

\begin{figure*}
	\centering
	\includegraphics[width=8.5cm]{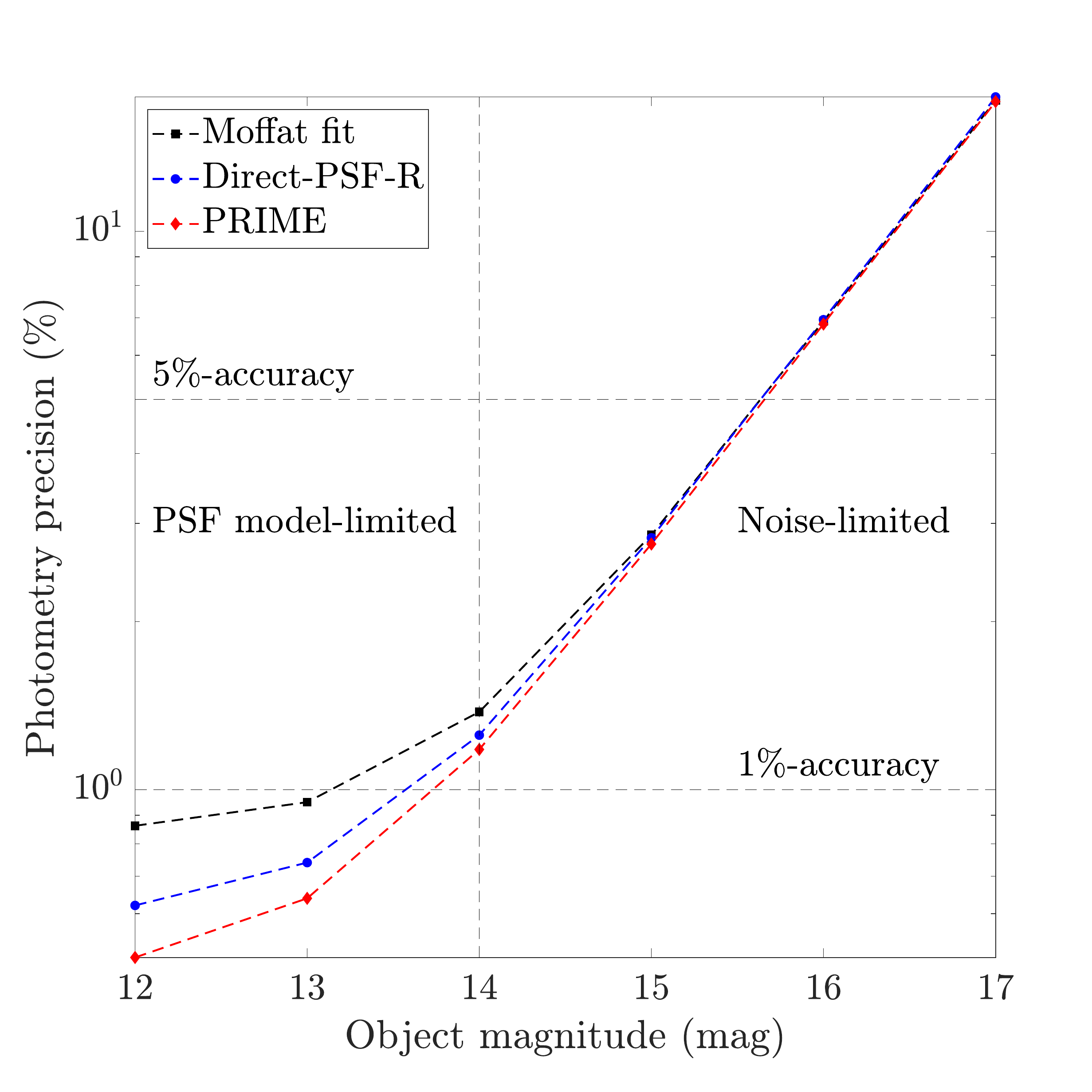}
	\includegraphics[width=8.5cm]{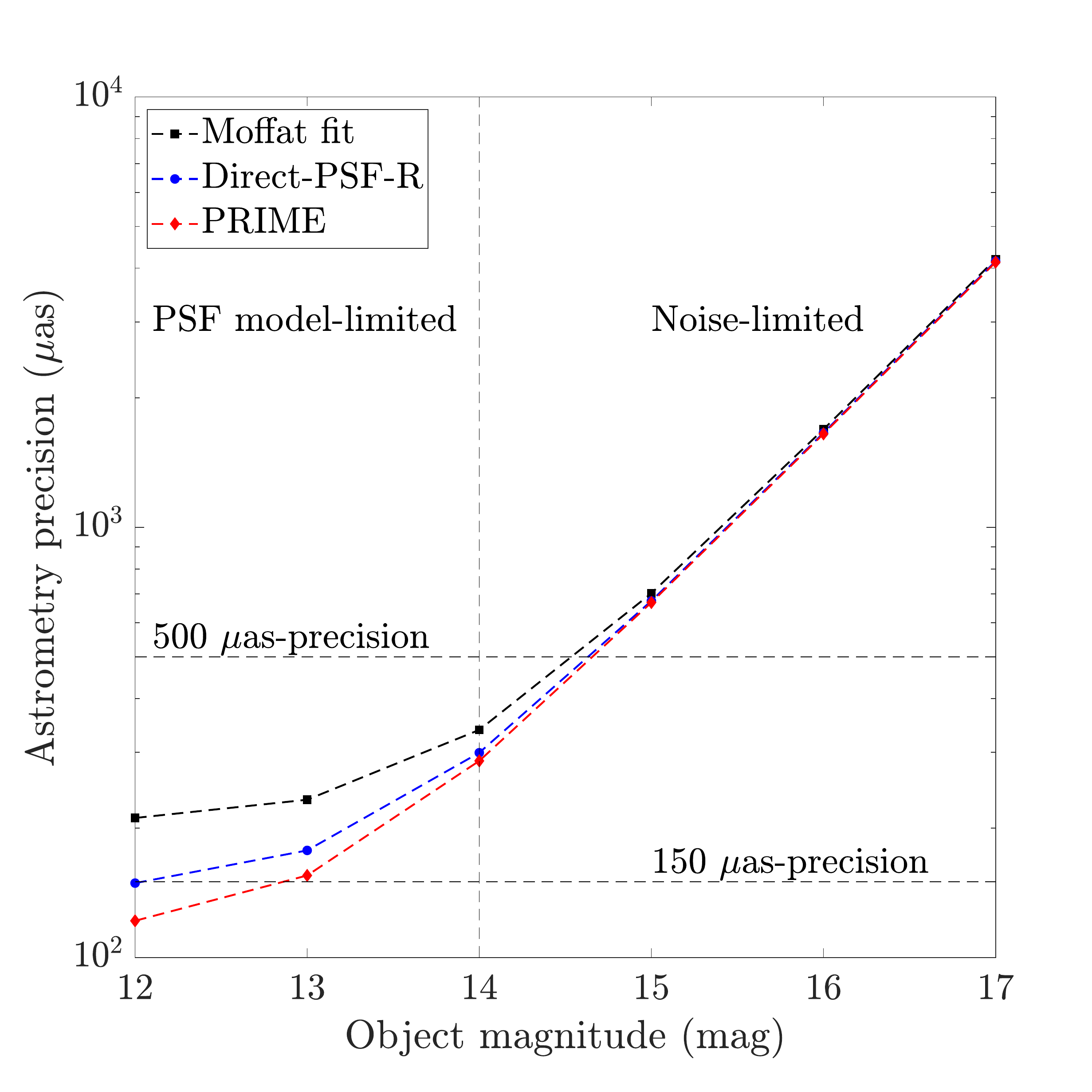}
	\caption{\textbf{Top~:} Photometry \textbf{Bottom~:} astrometry precision on binary object regarding the object magnitude (50s of exposure time) and the PSF model. Results are averaged out over the 5 frames of each of the two observed binary. }
	\label{F:photoBinaryVmag}
\end{figure*}

We illustrate in Fig.~\ref{F:photoBinaryVmag} the photometry and astrometry precision with respect to the artificial object magnitude. This plot shows two distinct regimes below and beyond magnitude 14 mag that respectively correspond to PSF-model and noise limitation regimes. For stars with $m_H > 14$ mag, the noise propagation through the fitting process dominates the estimation precision, while for stars with $m_H \leq 14$ mag, we clearly see the huge gain obtained thanks to PRIME. In this PSF-model limited regime, for a given object magnitude, we improve the precision by a factor 2 and 1.5 on respectively the photometry and astrometry; or in other words, we get the same precision for objects one to two magnitudes fainter.
Also, because PRIME does estimate PSF parameters in addition to photometry and astrometry, we were expecting more sensitivity to noise and see the metrics precision getting worse faster than the classical PSF-model, which does not occur.  Finally, we understand that the present implementation of PRIME is efficient when feeding it with image of stars of magnitude $m_H \leq 14$ mag stars with 50s of exposure time, which is accessible in the Galactic center (\cite{Yelda2010}) and makes it feasible to deploy PRIME on such a science case. In practice, we can provide several pieces of the field to PRIME to make sure we gather enough S/N for the PSF calibration. 

\subsection{Astrometry and photometry accuracy versus PSF model}

We have applied PRIME on 82 observations taken with NGS AO on 2017 Mar 15 of the triple system Gl 569 formed by a low-mass binary component Gl 569B located 4.9 arcsec from the primary Gl 569A. The AO loop was closed on Gl 569A and unsaturated images of all three components were obtained in the K cont filter ($\lambda_\text{eff} =2.2705 \mu$m) over 20 coadds of 0.2s exposure each.

Thanks to the presence of the third companion Gl 569A, we have extracted out the PSF from the scientific image and use it to estimate reference values for characterizing the separation and differential photometry of Gl 569B by using the non-linear minimization process implemented into PRIME. In other words, PRIME has solves the criterion given in Eq.~\ref{E:criteria} by playing on $[p_1,p_2]$ and $[\alpha_1,\alpha_2]$ with $\wOTF$ given by the 2D Fourier transform of the extracted on-axis PSF. 

Furthermore, we have played the exact same game by providing to PRIME three different PSF models. Firstly, we have predicted the PSF from the AO control loop data and the predictive laws for parameters proposed in Sect.~\ref{SS:prediction}. Secondly, we have best-fitted a asymmetric Moffat model on the Gl 569A image and use it as our PSF calibration. Finally, we have calibrated the PSF model described in Sect.~\ref{SS:theory} on the Gl 569A image by retrieving seeing, optical gains. For the three situation, we have also convolve the on-axis PSF model to a angular anisoplanatism filter obtained from MASS/DIMM $\cnh$ values.

\begin{figure*}
	\centering
	\includegraphics[width=17cm]{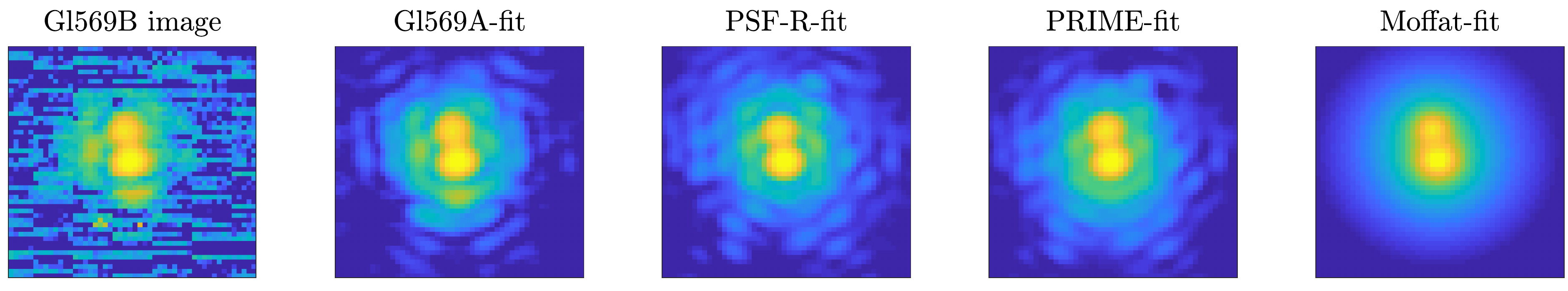}	
	\caption{\textbf{From left to right} \textbf{1)} H-band NIRC2 image of Gl589B acquired over 20 coadds of 0.2s integration each \textbf{2)} Best-fit using the the on-axis NGS image \textbf{3)} Best-fit using the PSF-R model \textbf{4)} Best-fit using the myopic PRIME estimation \textbf{5)} Best-fit using the myopic estimation based on a Moffat PSF model.}
	\label{F:Gl569B}
\end{figure*}

We report in Fig.~\ref{F:Gl569B} results of the fitting process for a particular acquisition compared to the original image of Gl 569B. The figures highlight that the binary observation is very well reproduced when relying on the extracted PSF; regarding the position of Gl 569A to the binary and the imaging wavelength, the anisoplanatism effect is extremely weak and can be neglected, in a way that the PSF extraction does not suffer substantially from any PSF model error. The reconstruction stands very close to the real PSF but miss its asymmetric shape, which is certainly introduced by a particular pattern of co-phasing errors of the primary mirror segments that produces low-order modes aberrations (\cite{Ragland2018_COPHASING}). PRIME succeeds partially restoring the asymmetric shape while the Moffat model is not accurate enough to mimic the complex PSF structure.

We report in Tab.~\ref{T:stats2017} estimation results averaged over the 82 acquisitions which confirm the visual inspection on Fig.~\ref{F:Gl569B}: PRIME allows to catch up reconstruction errors to achieve stellar parameters retrieval similar to the PSF extraction approach technique. It estimates $66.73" \pm 1.02$ of separation compared to the reference value of $66.76" \pm 0.94$ and $0.538$ mag $\pm 0.048$ of differential photometry compared to $0.532$ mag $\pm 0.041$ obtained from the exact PSF model. We were expecting the prediction process to mitigate the photometric bias as observed in Fig.~\ref{F:magErrorVseeing}, but the PSF was perturbed by co-phasing errors, which we can not predict so far, and strong telescope oscillations outside the regime of observing conditions of the test data sets, which can be mitigated by extending the number of test samples. PRIME shows already excellent agreements with the extracted PSF which gives evidence that PSF-R can be coupled efficiently with PSF extraction tools.

Also, we have played the game of estimating directly the PSF and stellar parameters on the binary image itself by using PRIME. We reached very same results, differences are within the uncertainty bars, although Gl 569B is 3.7 mag fainter than Gl 569A, which suggests that the joint estimation should be more efficient to the sole PSF calibration at equivalent S/N and points out that providing multiple sources PSF to PRIME allows effectively to strengthen the best-fitting process and mitigate the noise contamination. 
\begin{table}
	\centering
	\caption{Statistics on estimation of the binary separation and differential photometry regarding the PSF model. Error bars are obtained from an average over 82 successive acquisitions.}	
	\begin{tabular}{|c|c|c|}
		\hline
		PSF model & Separation (mas) & $m_A-m_B$ (mag)  \\		
		\hline
		Gl 569A & 66.76 +/- 0.94 & 0.532 +/- 0.041\\		
		\hline
		PRIME &  66.73 +/- 1.02& 0.538 +/- 0.048\\ 
		\hline
		PSF-R  &64.50 +/- 1.00 & 0.628 +/- 0.051 \\
		\hline
		Moffat &68.17 +/- 1.35 &  0.689 +/- 0.061\\ 
		\hline
	\end{tabular}
	\label{T:stats2017}
\end{table}

\begin{figure*}
	\centering
	\includegraphics[width=8.5cm]{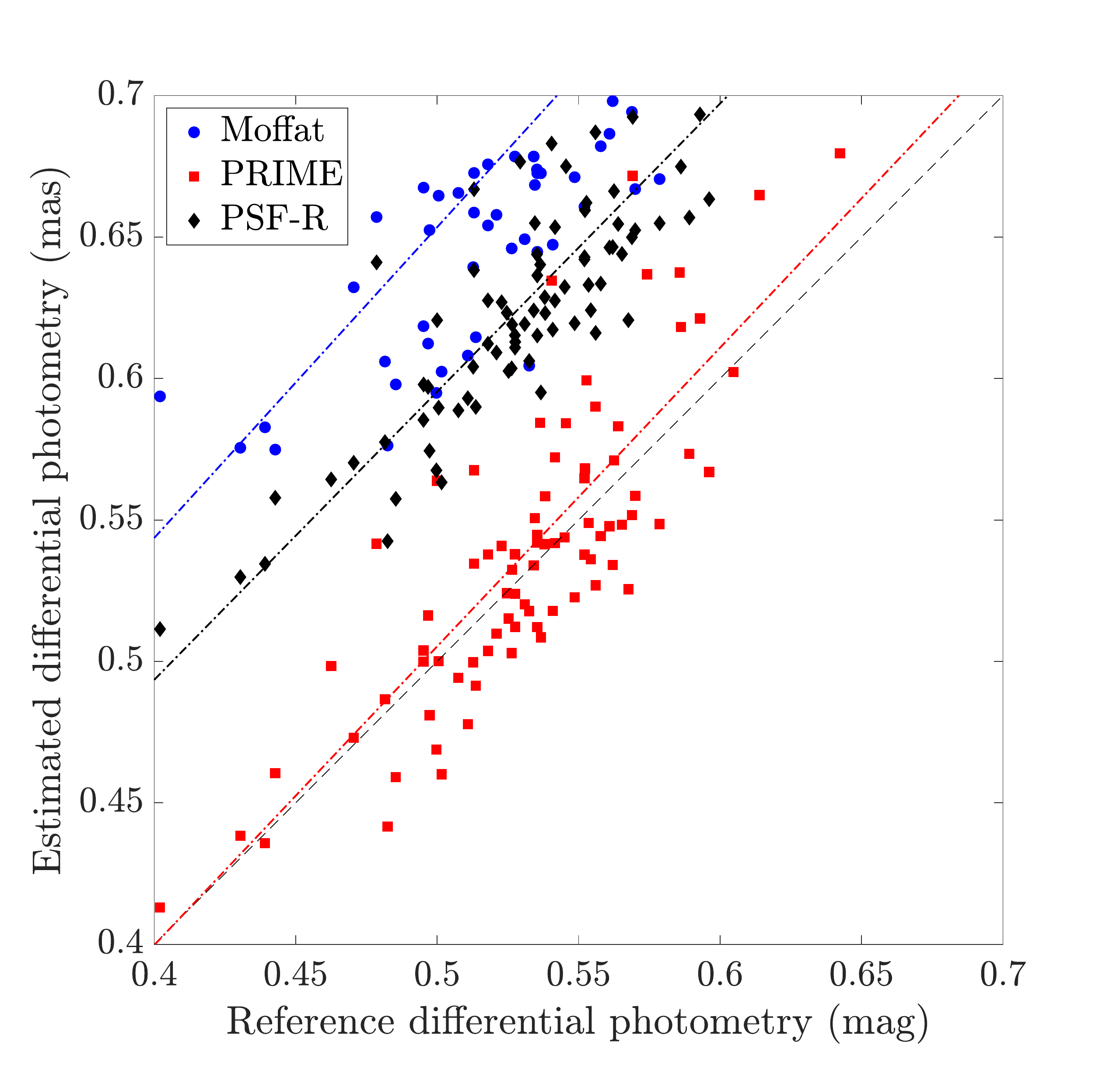}	
	\includegraphics[width=8.5cm]{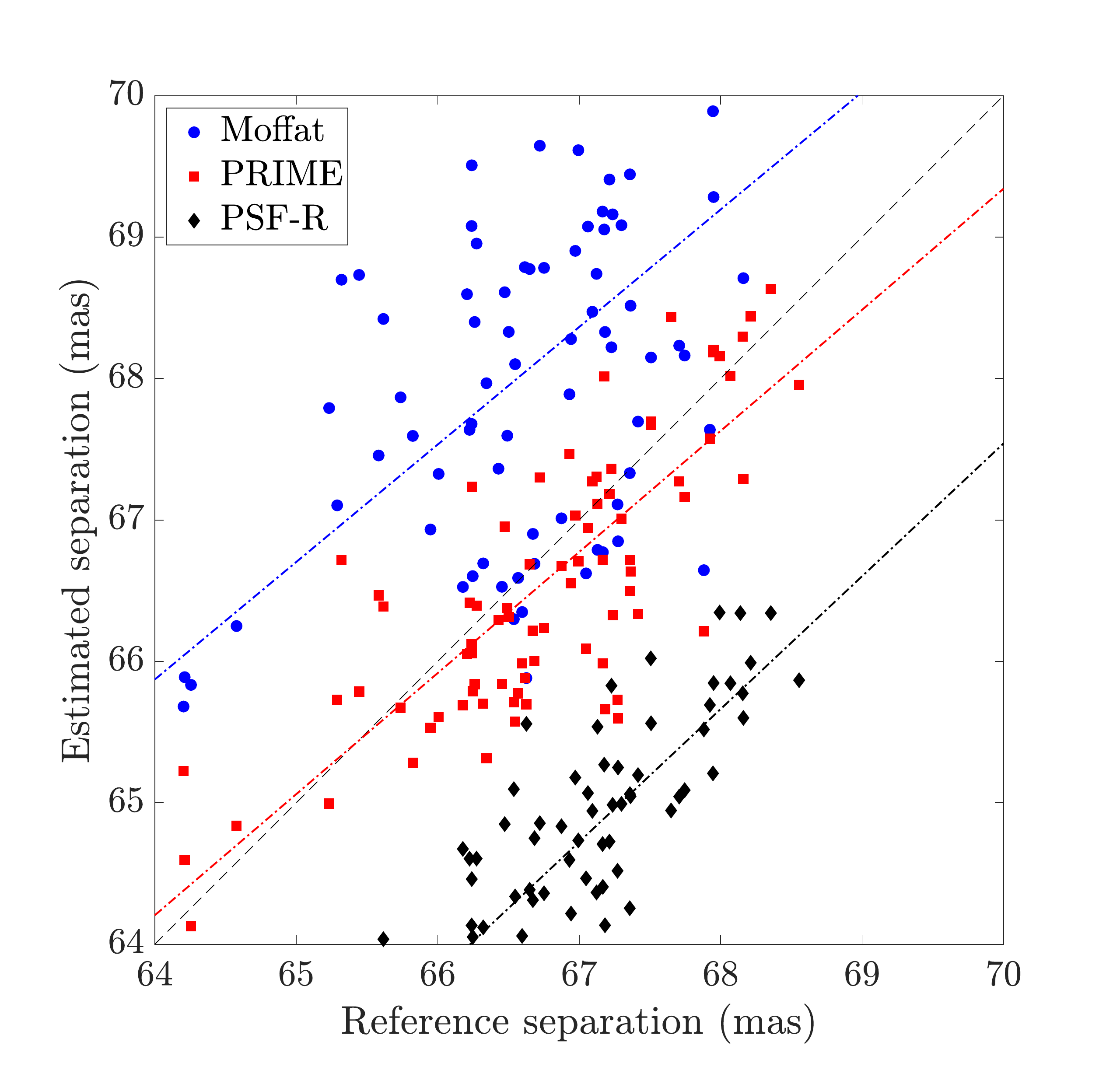}
	\caption{ Estimated \textbf{Top~:} differential photometry \textbf{Bottom~:} binary separation  obtained by modeling the PSF with a Moffat, PSF-R and PRIME with respect to the Gl569A-based retrieval. Each sample corresponds to one of the 82 acquisitions. The dash line corresponds to y=x and the dashed-dotted lines refer to linear regression for each model. }
	\label{F:binaccuracy}
\end{figure*}

\section{Conclusion}
\label{S:conclusion}
	
We have introduced PRIME as a new approach that couples AO telemetry and focal plane images to provide an accurate PSF model across field locations and wavelengths. PRIME has the ability to calibrate the atmospheric and system parameters that the AO telemetry-based PSF model depends on, such as the $\cnh$ profile and WFS optical gains, using a non-linear least-square minimization routine that allows to mitigate temporal variations and poor-knowledge of model inputs. In addition to providing the PSF model, PRIME yields an assessment of those input variations regarding the observation conditions and allows to scale up the wavefront error breakdown absolute value with respect to both the AO telemetry and the image.\\

We have deployed PRIME on Keck II on-sky PSF in engineering mode when guiding the system either on a NGS or a LGS on-axis. We have validated that only few parameters must be estimated carefully (seeing, optical gains plus additional focus and astigmatism terms) to reach -0.7\% $\pm$ 0.27 of accuracy on SR, 3.2 mas $\pm$ 1.2 on FWHM and 0.001 $\pm$ 0.004 on photometry. Over a sub-sample of data, we have identified empirical laws of these parameters with respect to the DIMM seeing that have been used to predict the PSF from the direct model. We have shown that this calibration permits to increase the absolute photometry accuracy from 0.22 mag $\pm$ 0.12 down to -0.03 mag $\pm$ 0.06.  \\

We have deployed PRIME over tight-binaries observations and analyzed, on the one hand the robustness to noise, on the other hand the PSF model accuracy. We have shown on NIRC2 binary images that PRIME is sufficiently robust to noise to retain photometry and astrometry precision below 0.005 mag and 100$\mu$as on a $m_H=14$ mag object. Finally, we have also validated that PRIME can perform a PSF calibration on the triple system Gl569BAB which provides a separation of 66.73$\pm 1.02$ and a differential photometry of 0.538$\pm 0.048$, compared to the reference values obtained with the extracted PSF which are 66.76 mas $\pm$ 0.94 and 0.532 mag $\pm$ 0.041.\\

PRIME is capable of calibrating the PSF model in the case of observations of the globular clusters and future application to engineering data will permit to make the PSF prediction more robust and efficient. The next step of this work will consist in coupling existing sources extraction softwares with PRIME to increase the measurements accuracy of astrometry and photometry on crowded fields observations. Moreover, we will deploy the PSF-based AO diagnostic to improve \textit{a posteriori} the AO running operations. The algorithm is currently implemented in Matlab and will be migrated into Python and made accessible to the community.

\section*{Acknowledgments}
The research leading to these results received the support of the
A*MIDEX project (no. ANR-11-IDEX-0001-02) funded by the ”Investissements
d'Avenir” French Government program, managed by the French National
Research Agency (ANR). This work has received partial funding from the European Union’s Horizon 2020 research and innovation programme under grant agreement No 730890. This work was also supported by the Action Spécifique Haute Résolution Angulaire (ASHRA) of CNRS/INSU co-funded by CNES.

This material reflects only the authors views and the Commission is not liable for any use that may be made of the information contained therein.
The data presented herein were obtained at the W. M. Keck Observatory, which is operated as a scientific partnership among the California Institute of Technology, the University of California and the National Aeronautics and Space Administration. The Observatory was made possible by the generous financial support of the W. M. Keck Foundation. The authors wish to recognize and acknowledge the very significant cultural role and reverence that the summit of Maunakea has always had within the indigenous Hawaiian community.  We are most fortunate to have the opportunity to conduct observations from this mountain.   
	
\bibliographystyle{plain} 
\bibliography{/home/omartin/Documents/Bibliography/biblioLolo}

\end{document}